\documentclass[fleqn,usenatbib]{mnras}

\usepackage{newtxtext,newtxmath}

\usepackage[T1]{fontenc}

\DeclareRobustCommand{\VAN}[3]{#2}
\let\VANthebibliography\thebibliography
\def\thebibliography{\DeclareRobustCommand{\VAN}[3]{##3}\VANthebibliography}

\usepackage{graphicx}	
\usepackage{amsmath}	
\usepackage{CJKutf8}    
\usepackage{color,soul} 
\usepackage{verbatim}   
\usepackage{pdflscape}  

\usepackage{tikz,xcolor,hyperref}

\newcommand{\mgii}{Mg\,\textsc{ii}\ }
\newcommand{\civ}{C\,\textsc{iv}\ }
\newcommand{\mgiins}{Mg\,\textsc{ii}}
\newcommand{\civns}{C\,\textsc{iv}}
\newcommand{\feii}{Fe\,\textsc{ii}}

\newcommand{\hbeta}{H\textsc{$\beta$}\ }
\newcommand{\hbetans}{H\textsc{$\beta$}}

\newcommand{\civl}{C \textsc{iv} $\lambda$1549.06}
\newcommand{\feiil}{Fe \textsc{ii} $\lambda$4570}

\newcommand{\civmgii}{\civns/\mgiins}

\newcommand{\fwhb}{FWHM(\hbetans)}
\newcommand{\fwcm}{FWHM(\civmgii)}
\newcommand{\dvciv}{$\Delta v$(\civns)}
\newcommand{\dvcm}{$\Delta v$(\civns -\mgiins)}
\newcommand{\aoff}{$\Delta \log A_0$}
\newcommand{\loff}{$\Delta \log L_{3000}$}
\newcommand{\lrf}{$\lambda_{\rm rf}$}
\newcommand{\tth}{$t_{\rm th}$}
\newcommand{\dttth}{$\Delta t$/\tth}
\newcommand{\rblr}{$R_{\rm BLR}$}
\newcommand{\vlos}{$v_{\rm los}$}
\newcommand{\vloc}{$v_{\rm local}$}
\newcommand{\bloc}{$\beta_{\rm local}$}
\newcommand{\bmin}{$\beta_{\rm local,min}$}
\newcommand{\bmax}{$\beta_{\rm local,max}$}

\definecolor{lime}{HTML}{A6CE39}
\DeclareRobustCommand{\orcidicon}{
    \begin{tikzpicture}
    \draw[lime, fill=lime] (0,0) 
    circle [radius=0.16] 
    node[white] {{\fontfamily{qag}\selectfont \tiny ID}};
    \draw[white, fill=white] (-0.0625,0.095) 
    circle [radius=0.007];
    \end{tikzpicture}
    \hspace{-2mm}
}

\newcommand{\orcidJJT}{\href{https://orcid.org/0000-0002-1860-0886}{\orcidicon}}
\newcommand{\orcidChrisW}{\href{https://orcid.org/0000-0002-4569-016X}{\orcidicon}}
\newcommand{\orcidJT}{\href{https://orcid.org/0000-0003-2858-9657}{\orcidicon}}
\newcommand{\orcidSamuel}{\href{https://orcid.org/0000-0001-9372-4611}{\orcidicon}}
\newcommand{\orcidSukYeeY}{\href{https://orcid.org/0000-0002-5204-2902}{\orcidicon}}
\newcommand{\orcidZac}{\href{https://orcid.org/0009-0007-5201-0357}{\orcidicon}}

\title[Quasar viewing angle and variability]{Probing Quasar Viewing Angle with the Variability Structure Function}

\author[J.-J. Tang et al.]
{Ji-Jia Tang$^{1,2}$\orcidJJT \thanks{E-mail: ji-jia.tang@anu.edu.au, ji-jia.tang@phys.ntu.edu.tw},
Christian Wolf$^{1,3}$\orcidChrisW , 
John Tonry$^4$\orcidJT , Samuel Lai$^1$\orcidSamuel , Suk Yee Yong$^{5,6,7,8}$\orcidSukYeeY and
\newauthor
Zachary Steyn$^1$\orcidZac \\
$^1$Research School of Astronomy and Astrophysics, Australian National University, Cotter Road, Weston Creek ACT 2611, Australia \\
$^2$Graduate Institute of Astrophysics and Department of Physics, National Taiwan University, No. 1, Sec. 4 Roosevelt Road, Taipei 10617, Taiwan \\
$^3$Centre for Gravitational Astrophysics (CGA), Australian National University, Building 38 Science Road, Acton ACT 2601, Australia \\
$^4$Institute for Astronomy, University of Hawaii, 2680 Woodlawn Drive, Honolulu, HI 96822-1897, U.S.A. \\
$^5$ARC Centre of Excellence for All Sky Astrophysics in 3 Dimensions (ASTRO 3D) \\
$^6$Astrophysics and Space Technologies Research Centre, Macquarie University, Sydney, NSW 2109, Australia \\
$^7$School of Mathematical and Physical Sciences, Macquarie University, Sydney, NSW 2109, Australia \\
$^8$Australian Astronomical Optics (AAO), Faculty of Science and Engineering, Macquarie University, Sydney, NSW 2109, Australia \\
}

\date{Accepted XXX. Received YYY; in original form ZZZ}

\pubyear{2023}

\begin{document}
\label{firstpage}
\pagerange{\pageref{firstpage}--\pageref{lastpage}}
\maketitle

\begin{abstract}
Given the anisotropic emission from quasar accretion discs, their viewing angle affects estimates of the quasar luminosity, black-hole mass and Eddington ratio. Discs appear overluminous when viewed pole-on and underluminous when viewed at high inclination. In radio-quiet quasars, the viewing angle is usually unknown, although spectroscopic indicators have been proposed. Here, we use a recently discovered universality in the variability structure function (SF) of quasar light curves (LCs), where all quasars show the same SF when clocks run in units of orbital timescale. As an offset from the mean relation can be caused by incorrect orbital timescales and thus incorrect luminosities, we correlate these offsets with suggested inclination indicators. We derive SFs from NASA/ATLAS LCs spanning $\sim 6$ years of observation, using a sample of 183 luminous quasars with measured \hbeta lines as well as 753 quasars with \civ and \mgii lines. Starting from the proposed orientation indicators, we expect quasars with narrower \hbeta lines and with more blueshifted \civ lines to be viewed more pole-on and thus appear overluminous. In contrast, our SF analysis finds that presumed pole-on discs appear underluminous, consistently for both line indicators. We discuss possible explanations for the behaviour of quasars with highly blueshifted \civ lines irrespective of inclination angle, including dusty outflows that might render the accretion disc underluminous and flatter disc temperature profiles with longer orbital timescales than in thin-disc models but reach no satisfying conclusion.
\end{abstract}




\defcitealias{TWT}{Paper I}

\section{Introduction}\label{intro}

The strongest sources of UV-optical light in the Universe are fast-growing supermassive black holes (SMBH), appearing as quasars. These can shine up to $\sim$1\,000 times brighter than galaxies and can thus be observed with relative ease into great distances \citep[e.g.][]{Schm70, Rich06, Wolf18, On23}, all the way into the early Universe. Two open questions on quasars have been lingering for a long time now: (1) What is the nature and geometry of powerful outflows observed in blueshifted emission lines? And (2) are there ways of gauging the viewing angle to the symmetry axis of the black-hole environment? Potentially, if the outflows are ubiquitous but not isotropic, they could offer a diagnostic for the viewing angle \citep[e.g.][]{Bald97, Rich11, Yang20}.

The unusually high luminosity of quasars is powered by release of gravitational energy in the potential well of a black hole \citep{Lyn69} and friction that infalling matter experiences in an accretion disc \citep{SS73}. This friction not only slows down the matter orbiting the black hole by enough to make it fall in and not escape; it also converts the gravitational energy of the matter into heat. The accretion disc is believed to be optically thick, meaning it is not transparent to light, while it is a source of most intense black-body radiation itself. Crucially, the disc is geometrically thin, meaning it is rendered almost as a 2D surface. Hence, the orientation of a disc relative to the line-of-sight to an observer affects the observed brightness \citep{NT73, Li05}, which also contributes to the global Baldwin effect \citep{Net85}.

The orientation of a quasar accretion disc, or viewing angle it presents, matters not only for our perception and understanding of quasars, but also limits what they can tell us about the Universe. Our estimates of the luminosity for a quasar are affected by the viewing angle to its accretion disc, which is usually unknown. The luminosity translates into a mass accretion or growth rate for the black hole, which is equally affected by the viewing angle. The problem also propagates into our estimate of a quasar's output of ionising radiation \citep{Fan06}. From that, we further estimate the size of the broad emission line region (BLR) and hence the mass of the black hole, at least in the most common "single-epoch virial mass measurements", which are both subject to viewing angle effects \citep[e.g.][]{Ves02, SL12}. Even more elaborate methods for measuring black hole masses such as reverberation mapping (RM) still depend on viewing angles and emission geometry for their results \citep[e.g.][]{NP97, Pet04, Du14, Panco14}. Viewing angle has other effects too: the wavelength of spectral lines in quasars reflects the radial velocities of the line-emitting gas. Hence, the orientation effect for the line-width needs to be considered in the virial factor and as a consequence the black hole mass estimation depends on it \citep[e.g.][]{MR18}. An unknown viewing angle thus makes it harder to infer the true physical speeds in Keplerian orbital motion and in outflows. 

Finally, quasars are only one manifestation in the unified scheme of Active Galactic Nuclei \citep[AGN;][]{UP95}. The nucleus is assumed to include dust in a torus-like region that is aligned with the plane of the accretion disc. In that scheme, an observer would see accretion discs obscured by the dusty torus when viewing them nearly edge-on, rendering those growing SMBHs as "type-2 AGN" in contrast to quasars ("type-1 AGN"). Studies of the obscured fraction of AGN suggest that roughly half of them are viewed through a dusty torus if that is the cause of the type-2 appearance, and that this fraction declines towards higher accretion luminosity \citep[e.g.][]{Hasinger08, Ass13}, an effect known as receding torus \citep{Law91} although other interpretations are possible \citep[see discussion in][]{Pad17}. Assuming a random orientation of the parent AGN sample, a 50\% fraction suggests that quasars are obscured at inclinations from $i=90^\circ$ (edge-on) to $i\approx 60^\circ$ as that range covers half the solid angle seen by the accretion disc. However, this statistical inference assumes a torus without holes, and measuring disc inclinations in quasars, i.e. unobscured AGN, could shed light on the true dust geometry, which may also depend on the radiative power of the quasar. 

While the viewing angle of a quasar is arguably important in many ways then, it has proven very difficult to measure. So far, the accretion disc of a quasar has appeared unresolved in any UV-optical-IR imaging, as the emitting disc is smaller than resolution limits of existing telescopes. Owing to its luminosity that outshines entire galaxies, the disc has even rendered the quasar host galaxies almost unobservable \citep[e.g.][]{Disney95, Dun03}. To overcome this barrier, two ways forward are explored: (1) using secondary indicators, whose orientation is linked to that of the disc and which can be resolved; and (2) using indicators that work with the unresolved nuclear signal itself.

One particular method, BLR Tomography, has been used to reconstruct the structure and orientation of the disc-like BLR from intense time-domain monitoring of Balmer lines \citep[e.g.][]{Panco14, Bentz22}. Variations in line appearance per radial velocity channel allow such reconstructions, but require observations over a period of time and assume that the geometry of the BLR does not change during the observations. Such observations have produced impressive model pictures of the BLR for a handful of nearby Seyfert AGN, but the required timescales for observing quasars would be challenging. 

A milestone of interferometric imaging is the observation of the disc-like Pa-$\alpha$ BLR in the iconic quasar 3C 273 by the Very Large Telescope (VLT) GRAVITY Collaboration \citep{gravity18}. Spatial resolution still limits this method to the quasars with the largest angular BLR extent, which scales roughly with apparent magnitude. A forthcoming GRAVITY+ instrument upgrade will improve sensitivity but not resolution, and hence not help with resolving by far most quasars in the population. 

The NLR in host galaxies extends to kpc scales and represents nebular ionisation cones bounded by the opening angle of the dusty torus in the nucleus \citep[e.g.][]{UP95, DM18}. These can be observed in nearby low-luminosity Seyfert AGN \citep{Thom17} and obscured quasars \citep{ZG14}, but are difficult to map in unobscured quasars that outshine their host galaxies \citep[e.g.][]{Huse16}.

Radio-loud AGN constitute only a $\sim$10 -- 20\% fraction of AGN \citep[e.g.][]{Jiang07, Ban15}. In some of these, relativistic jets are visible in the nuclear region, where interferometric radio observations can see the motions of jet blobs and constrain the viewing angle \citep[e.g.][]{Pear81, Ghis93, An10}. The largest-scale products of these bipolar jet outflows are radio lobes, which can extend to a Mpc scale and appear bi-polar only if not viewed pole-on \citep[e.g.][]{Per84}. \citet{Kim11} presented a valuable discussion of the correlations between line strengths and radio-to-optical flux ratios and their implications for BLR properties and their orientation effects. However, it is not clear to what extent this applies to the majority of quasars, which are radio-quiet.

Arguably an impressive avenue are observations of black hole shadows as accomplished by the Event Horizon Telescope with the SMBH in the galaxy M87 \citep{EHT19} and at the centre of the Milky Way \citep{EHT22}. As these observations see the innermost accretion region outside the photon sphere of the black hole, they can constrain the viewing angle of the accretion disc. However, such measurements have not yet been extended to distances where quasars are found.

While the above methods are intriguing and impressive, none of them seem able to address the bulk of the quasar population for a long time to come. This leaves us to explore ways of extracting viewing angle constraints from the unresolved nuclear signal itself. The nuclear signal includes emission from the BLR as well as blackbody emission from the accretion disc over a range of hot temperatures and from the dusty torus over a range of cooler temperatures, with a possibility of the dusty torus shielding part of the accretion disc \citep[see, e.g.][and discussion therein]{Temp21}.

Two widely discussed proxies for orientation in quasars relate to the appearance of broad emission lines. It has long been suggested that the BLR traces two types of motions among line-emitting gas clouds that are close to the accretion disc \citep[see e.g.][and discussion therein]{Bald97,Gas09}: 
\begin{enumerate}
\item Low-ionisation lines such as \mgii and \hbeta lines are believed to have a disc-like appearance coupled with an orbital motion close to the plane of the accretion disc \citep{Marz01, Marz22, Panco14, Run14, SH14, MR18}. Lines emitted from a rotating disc of clouds will appear largely symmetric and centred on the systemic velocity irrespective of orientation. The rotation will produce line widths that change with viewing angle and appear broadest closer to edge-on views.
\item High-ionisation lines, of which \civ is the strongest one, are believed to arise from outflows launched by winds in the inner regions of the disc \citep{Rich11,Wang13,Yong20}. Usually, only one of the outflow cones will be visible since the accretion disc is not transparent. The radial velocity distribution seen in outflows will drive the line shape to be asymmetric and generally blueshifted; polar views of a face-on disc would then cause the highest blueshifts. However, \citet{Run14} argue that \civ blueshifts do not reflect orientation but intrinsic variety in AGN. A high-\civ blueshift may indicate a stronger wind from the accretion disc \citep{Lei04} and may correlate with a high accretion rate \citep{Temp23}. Or, as often with a dichotomy of explanations, both of them may contribute and be hard to disentangle. 
\end{enumerate}

In this work, we take advantage of a new calibration for distance-independent luminosity measurements of quasar accretion discs, which is based on the structure function (SF) of the UV-optical continuum variability in quasars. \citet{TWT} (hereafter \citetalias{TWT}) found that the variability SF is a universal law once the time intervals in the SF are expressed in units of the characteristic orbital or thermal timescale (\tth) of the accretion disc. When quasars of similar apparent magnitude and luminosity distance show different SFs, one interesting possibility is that their luminosity estimate is affected by dust extinction or by orientation effects. In the following, we analyse the light curves (LCs) from the NASA/ATLAS survey of nearly 1\,000 bright quasars with either \civ blueshift or \hbeta full-width half maximum (\fwhb{}) data from SDSS DR14 spectra \citep{Rak20}, and examine whether we find an inclination signature or instead signs of intrinsic variety among AGN.

In Section~\ref{sec:data_samp}, we describe the quasar sample and data used in this study. In Section~\ref{sec:VSF} we describe the construction of the variability SF and in Section~\ref{sec:acc-disc-lum} we describe how we derive the expected luminosity bias from models of accretion disc emission observed in a General Relativity framework. In Section~\ref{sec:mod} we present how we relate orientation to spectral line observables, and in Section~\ref{sec:results} we present our results and discussion. We adopt a flat $\Lambda$CDM cosmology with $\Omega_\Lambda = 0.7$ and $H_0 = 70$~km~s$^{-1}$~Mpc$^{-1}$. We use Vega magnitudes for data from {\it Gaia} and AB magnitudes for data from NASA/ATLAS and for absolute magnitude estimates.

\section{Data and sample}\label{sec:data_samp}

As in \citetalias{TWT}, we analyse the variability in the LCs of the brightest quasars in the sky. Thus, we start from an identical quasar sample as \citetalias{TWT}. We use the same LCs from NASA/ATLAS \citep[Asteroid Terrestrial-impact Last Alert System;][]{Tonry18} as \citetalias{TWT} did after the same cleaning steps were applied. Below, we first summarise the steps taken in \citetalias{TWT}, before we reduce the parent quasar sample to two specific subsamples with spectroscopic diagnostics that have been proposed as indicators of viewing angle.

\subsection{ATLAS data and parent sample}\label{sec:data}

\citetalias{TWT} combined the Million Quasars Catalogue \citep[MILLIQUAS v7.1 update;][]{Flesch15} with the {\it Gaia} eDR3 catalogue \citep{gaia21} to select 6\,163 spectroscopically confirmed, non-lensed\footnote{\href{https://research.ast.cam.ac.uk/lensedquasars/}{https://research.ast.cam.ac.uk/lensedquasars/}}\citep{Le19, Le23} quasars with $Gaia$ magnitude $Rp<17.5$, redshift $0.5<z<3.5$, declination $\delta>-45^\circ$ and Galactic foreground reddening \citep{SFD98} of $E(B-V)_{\rm SFD}<0.15$. They also selected quasars to be isolated by requiring a Gaia BpRp Excess Factor below 1.3. For these quasars, the NASA/ATLAS database provides LCs in two passbands, cyan (5330\AA) and orange (6785\AA), whose magnitudes are denoted by $m_{\rm c}$ and $m_{\rm o}$, respectively. \citetalias{TWT} then rejected quasars with noisy LCs, i.e., where the 90-percentile flux errors in the LCs were above $\sigma_{f_\nu, {\rm o}} > 150\mu$Jy or $\sigma_{f_\nu, {\rm c}} > 85\mu$Jy. 

They rejected radio-loud quasars, whose optical variability may have origins beyond the accretion disc. For that, they cross-matched the sample with catalogues from the Faint Images of the Radio Sky at Twenty-cm \citep[FIRST;][]{Be95} Survey, the NRAO VLA Sky Survey \citep[NVSS;][]{Co98}, and the Sydney University Molonglo Sky Survey \citep[SUMSS;][]{Ma03}. The matching radii between the Gaia coordinates and the FIRST, NVSS, and SUMSS coordinates were 3", 12", and 11", respectively, to account for the different spatial resolution of the radio surveys. When multiple radio sources in NVSS or SUMSS were matched to one quasar, the one with the closest separation was used. The radio loudness criterion for matches with NVSS and SUMSS magnitudes ($t_{\rm NVSS/SUMSS}$) was $0.4\times(m_{\rm o}-t_{\rm NVSS/SUMSS})=1.3$, below which radio-detected objects were labelled as radio-intermediate. All quasars with matches in FIRST but not in NVSS were radio-intermediate. After that, 5\,315 quasars were left over.

The NASA/ATLAS LCs used in \citetalias{TWT} stretch from 2015 to 2021. LCs were cleaned by excluding all observations with large errors of $\log(\sigma_{f_\nu, {\rm o}})>-3.94-0.12\times(m_{\rm o}-16.5)$ and $\log(\sigma_{f_\nu, {\rm c}})>-4.17-0.10\times(m_{\rm c}-16.5)$; and further removed outliers by comparing each measurement with other observations within $\pm$7 days and then using a $2\sigma$-clipping technique to reject spurious outliers. Observations are retained if they are isolated within $\pm$7 days. \citetalias{TWT} found that data at rest-frame wavelengths \lrf$>3000$\AA~do not follow the simple mean relation (see Section~\ref{sec:VSF}), and they also avoided wavelengths that were short enough to be contaminated by Ly$\alpha$ emission. After discarding LCs from affected passbands, they were left with 7\,084 LCs from 4\,724 quasars.

Here, we match these 4\,724 quasars with the catalogue of SDSS DR14 quasar properties from \citet{Rak20}, and find 3\,253 matches with a separation of $<$1", mostly because the SDSS sample does not cover as much sky as Gaia and MILLIQUAS. In contrast to \citetalias{TWT}, we adopt the luminosity at 3000\AA~($L_{3000}$) from \citet{Rak20} as the latter is an estimate obtained from spectral modelling that decomposes spectral flux into an accretion disc continuum, a Fe template and the emission lines. In this work, we aim at detecting subtle trends in accretion disc luminosity with quasar viewing angle; hence, we need to use the best-possible estimate of luminosity for the disc alone, especially if there is a risk that spectral composition itself might depend on the viewing angle. Therefore, we reduce the sample to those 2\,913 quasars with good quality measurements of $L_{3000}$, i.e., those with LOG\_L3000 > 0 and QUALITY\_L3000 = 0 in the catalogue. In the following, we select which of these quasars may be used in a rotating BLR sample and in an outflow sample, where translations from line diagnostics to viewing angle have been proposed. 

\begin{figure}
\includegraphics[width=\columnwidth]{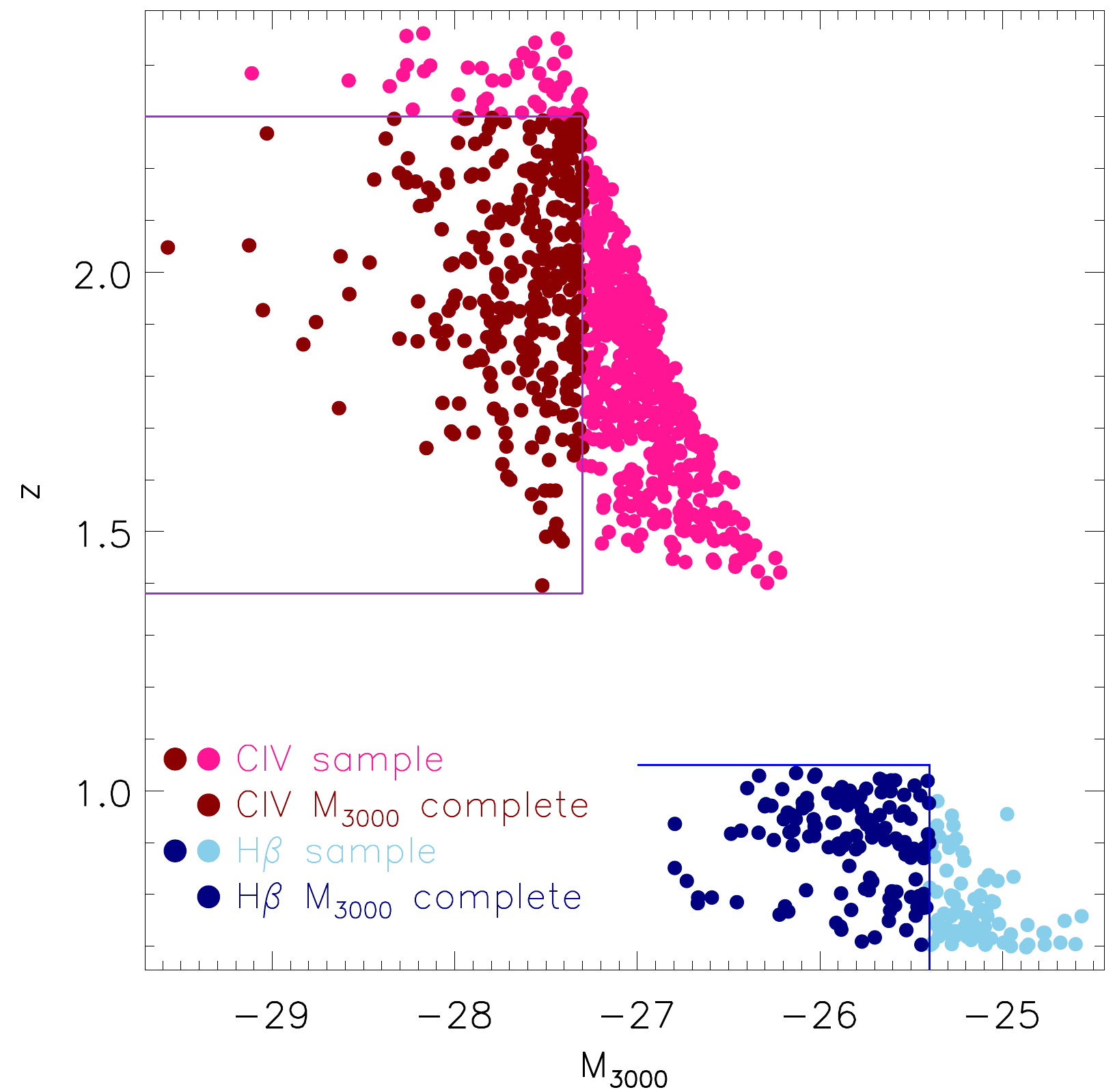}
\caption{Sample distribution of redshift $z$ vs. luminosity $M_{3000}$. Each of the two emission line samples relates to a different orientation indicator and has a full sample as well as a subsample that is complete in luminosity. \\
(A colour version of this figure is available in the online journal.)
}
\label{fig:mz}
\end{figure}

\subsection{Rotating BLR sample}\label{sec:hbsam}

We select a sample of quasars with FWHM measurements of the \hbeta line as an indicator of inclination assuming that the line originates in a rotating BLR \citep{Marz18}. From the \citet{Rak20} catalogue, we select non-broad absorption line (BAL) quasars (BAL\_FLAG < 1) with a median signal-to-noise ratio (SNR) per pixel of \hbeta (4750 -- 4950\AA) $\geq 15$, and thus find valid \hbeta peak and FWHM values for 313 quasars within our sample. Since we require rest-frame wavelengths of $\lambda <3000$\AA, only the cyan passband is used for the rotating BLR sample. 

\citet{Marz22} discuss in particular population xA quasars, which have flux ratios between the singly ionized iron emission blend centred at 4570\AA~and the broad component of \hbeta emission of $R_{\text \feii}=F({\text \feiil})/F({\text \hbetans})>1$; they argue that these tend to be highly accreting and have degeneracies between different inclination angles. Therefore, we exclude 104 population xA quasars, which we identify in our sample using the \citet{Rak20} catalogue.

Several studies showed that quasar variability amplitudes are suppressed on short timescales and show a break from the random walk regime \citep[e.g.][\citetalias{TWT}]{Mu11}. In this work, we aim to measure offsets between the average SF of quasar variability and that of individual quasars, which we will attempt to attribute to biased estimates of quasar luminosity. Hence, it is of great importance that we shall measure these offsets in the cleanly measured random walk regime of the SF. Breaks in the LCs thus restrict the range in the SF that is available for fitting offsets. Whenever such breaks are on timescales that appear to leave us with only a narrow fitting range, we discard the objects from the sample. Therefore, we remove 26 of the currently remaining quasars ($\sim 12$\%) and are left with 183 useful quasars in the full rotating BLR sample. We also select a subgroup of 104 quasars with absolute magnitude $M_{3000}<-25.4$ as a luminosity-complete rotating BLR sample for an alternative analysis. Figure~\ref{fig:mz} shows the $M_{3000}$-$z$ comparison of these two samples.

\subsection{Outflow sample}\label{sec:civsam}
We select a sample of quasars with measurements of the \civ line and the \mgii line as an indicator of inclination assuming that the \civ line originates in an outflow that will appear blueshifted relative to the \mgii line emitted from a rotating BLR much like the \hbeta line. From the \citet{Rak20} catalogue, we select non-BAL quasars (BAL\_FLAG < 1) with a median SNR per pixel $\geq 15$ in both \mgii (2700 -- 2900\AA) and \civ (1500 -- 1600\AA), which are the same criteria as used by \citet{Yong20}. We thus find valid peak wavelength and FWHM values for both lines for 873 quasars in our sample. 

\begin{figure}
\includegraphics[width=\columnwidth]{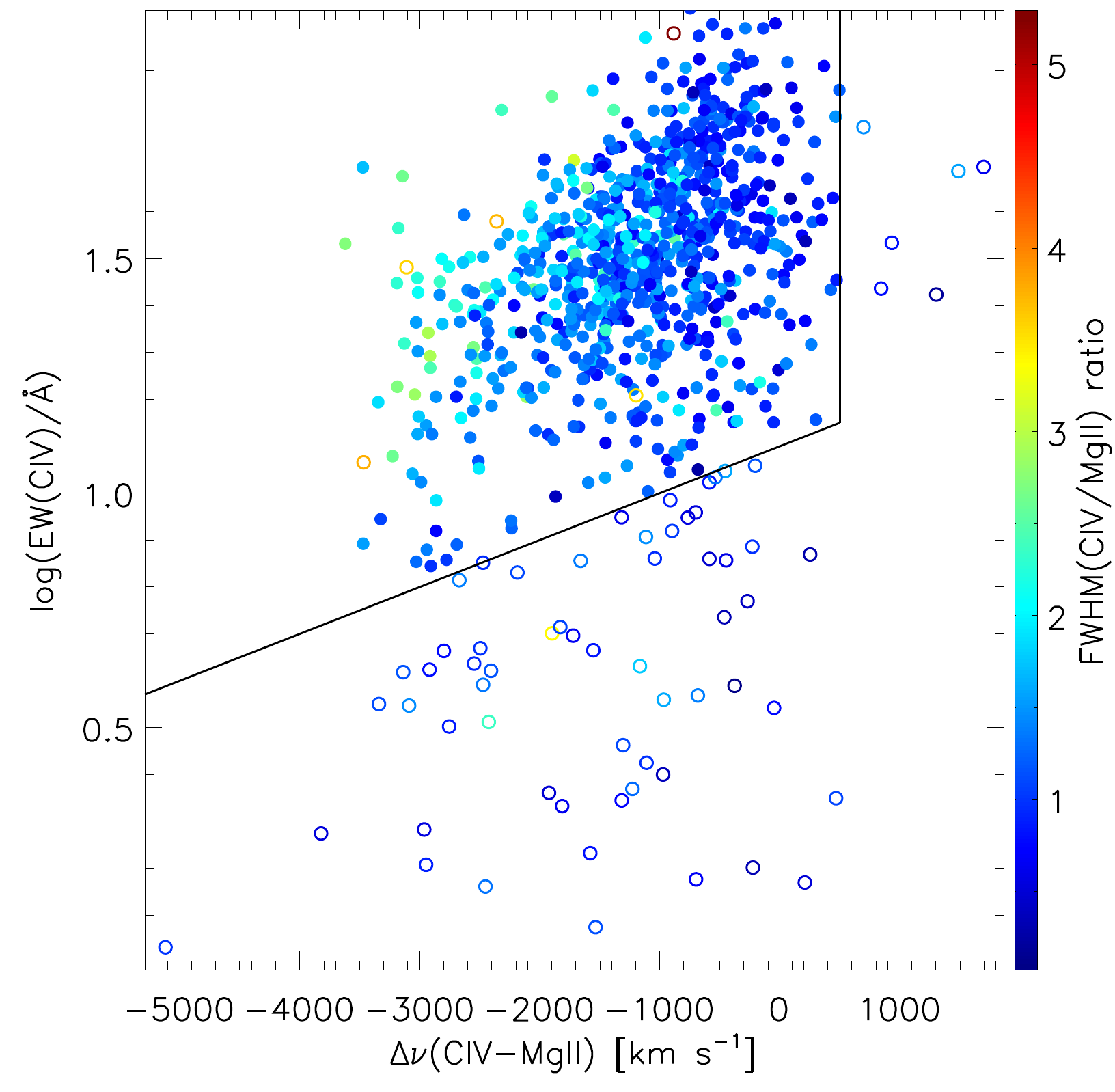}
\caption{Outlier removal process in the outflow sample: filled circles indicate the 803 selected vs. 70 rejected (open circles) quasars. Lines show the criteria for the two of the three selection axes, $\log$(EW(\civns)) and \dvcm. (A colour version of this figure is available in the online journal.)
}
\label{fig:civ_sam_sel}
\end{figure}

As shown in Figure~\ref{fig:civ_sam_sel}, we remove 70 outlier objects ($\sim 8$\%) that violate the following criteria: the \fwcm{} ratio $<3.5$, velocity shift \dvcm{} $<500$~km~s$^{-1}$, and \dvcm/(km~s$^{-1}) \times 0.0001+1.1 < \log $(EW(\civns)/\AA), where EW stands for equivalent width. Those objects with low-\civ EW line are noisy and unreliable measurements, thus they are removed. Similar to the \hbeta sample in Section~\ref{sec:hbsam}, we discard objects where breaks in the LCs make the fitting ranges too narrow for a robust offset determination (about 6\% of the sample). This gives us a total of 1\,464 SFs from 753 quasars in the full outflow sample. We also select a subgroup of 482 SFs from 256 quasars with redshift $z<2.3$ and $M_{3000}<-27.3$ as a luminosity-complete outflow sample (see also Figure~\ref{fig:mz}).

\begin{figure*}
\includegraphics[width=0.99\textwidth]{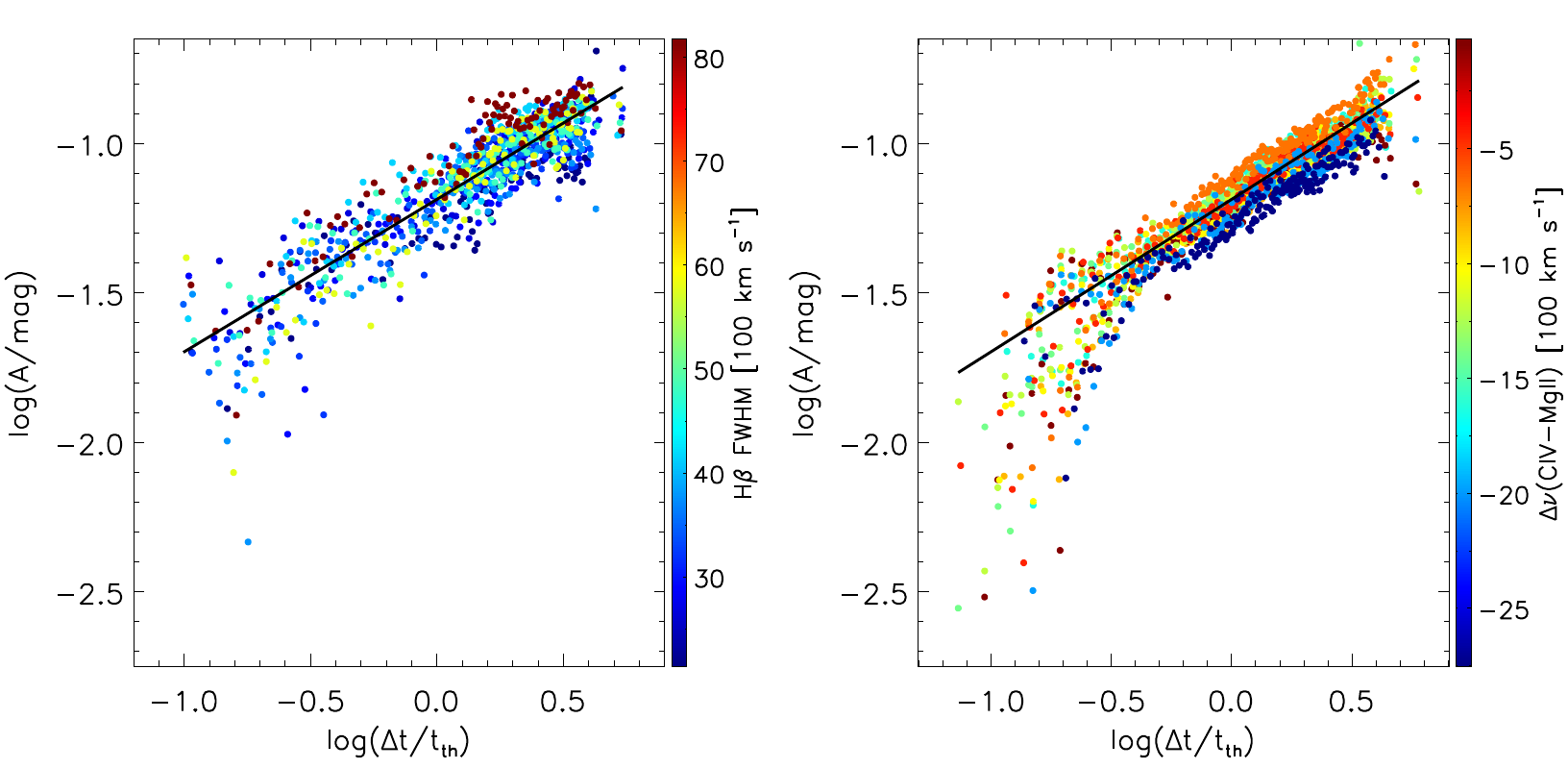}
\caption{Left panel: The structure function $A$(\dttth) from the full rotating BLR sample, which are binned and colour-coded according to their \fwhb. The black line is the mean relation (Equation~\ref{eq:A_ttherm}) in the random walk regime from \citetalias{TWT}.
Right panel: The SF $A$(\dttth) from the full outflow sample, which are binned and colour-coded according to their \dvcm. Note that deviations from the mean relation correlate with line properties. \\
(A colour version of this figure is available in the online journal.)
}
\label{fig:lumoff}
\end{figure*}

\section{Methods}\label{sec:method}

\subsection{Variability structure functions}\label{sec:VSF}

The structure function analysis has been developed to quantify the observed variability in a light curve, especially for the common cases of uneven sampling \citep{Hugh92}. The variability SF traces the measured difference between the brightness of an object at two different times as a function of time interval. \citet{Koz16} provides a substantial mathematical discussion of the SF analysis. In this work, we adopt the SF calculation from \citetalias{TWT}, which used the noise-corrected definition of variability amplitude from \citet{Cl96}:
\begin{equation}
	A=\sqrt{\frac{\pi}{2} <\Delta m>^2 - <\sigma^2>},   ~ ~ {\rm with} ~ ~
	\Delta m=|m_{\rm i}-m_{\rm j}|,
	\label{eq:sf}
\end{equation}
where $m_{\rm i}$ and $m_{\rm j}$ are any two observed apparent magnitudes and $\sigma$ is the magnitude error due to noise. The mean error $<\sigma^2>$ is determined as a function of observed magnitude, $m_{\rm obs}$, such that the SF for $\Delta t < 1$ day becomes $A\approx 0$. This approach is valid, because the typical amplitude of intraday variability in radio-quiet quasars is so small that it is within the uncertainty of the noise itself and will not noticeably affect the analysis of the random-walk portion in the SF. The required noise model for the given sample of ATLAS LCs was determined in \citetalias{TWT} and is given by 
\begin{equation}
	\log(<\sigma^2>)=n_0+n_1 m_{\rm obs} ~,
	\label{eq:noise}
\end{equation}
where $(n_0, n_1)=(-12.411, 0.585)$ and $(-12.428, 0.573)$ for the orange and cyan passband, respectively.

\citetalias{TWT} went beyond the traditional expression of the SF in units of rest-frame clock time and expressed them in units of the mean orbital or thermal timescale of the emission for any given quasar accretion disc and observing passband. This translation was based on the temperature profile in standard models of thin accretion discs \citep{SS73,FKR}. This translation of the time axis in the SF analysis removed the great diversity of SFs between quasars of different luminosity observed in different passbands. Instead, \citetalias{TWT} found a universal SF common to all quasars in their study and quantified the mean SF relation of the quasar variability in the random walk regime. Using \tth{}, the timescale for local disc components to reach thermal equilibrium in the standard thin disc models \citep{FKR}, and the scale length of the disc at given $L$, $\lambda$ values \citep{Mor10}, \citetalias{TWT} provided an equation to calculate the characteristic \tth{} for a quasar disc as
\begin{equation}
    \label{eq:tth}
	\frac{t_{\rm th}}{{\rm day}}=6.22\times10^{-27}\times(\frac{L_{3000}}{{\rm erg/s/\text{\AA}}})^{0.5}\times(\frac{1}{\cos{45^\circ}})^{0.5}\times(\frac{{\text \lrf}}{\text{\AA}})^2.
\end{equation}
For the calculation of \tth, they assumed that the observed $L_{3000}$ values of all quasars in the sample are affected by a disc inclination angle of 45$^\circ$, and applied a naive geometric correction factor ignoring relativistic effects. Any changes to this correction will simply rescale the units of the single \tth{} scale that is common to all quasars in the sample. They also took the cyan and orange pivotal wavelengths and quasar redshift to calculate \lrf{} and thus \tth{} for each quasar in each passband. Based on the above calculation, \citetalias{TWT} showed that in the random-walk regime the SF amplitude $A$ has a mean relation of
\begin{equation}
    \label{eq:A_ttherm}
    \log (A/A_0) = \gamma_{\rm th} \log ({\text \dttth}),
\end{equation}
where $\log A_0=-1.187\pm0.001$ and $\gamma_{\rm th}=0.510\pm0.002$. 

Although \citetalias{TWT} successfully discovered a universal relation and quantified the mean SF by changing to a thermal-timescaled clock, individual quasars still show modest offsets from the mean relation. In this work, we test the hypothesis that the amplitude offset is caused by an offset between the true luminosity and that measured by us. As we see quasar discs under a variety of inclination angles, these require a variety of correction factors to the non-isotropic luminosity $L_{3000}$ (see Sect.~\ref{sec:acc-disc-lum}) when attempting to infer the correct \tth. Here, we have assumed that \fwhb{} and \civ blueshift indicate the inclination angle. If instead they indicate a different structure of the accretion disc, then we will derive a wrong timescale by using the thin-disc approximation as in \citetalias{TWT} even when we observe the luminosity correctly.

We then determine amplitude offsets for individual quasars as well as mean offsets for quasar subsamples grouped by expected inclination. In the rotating BLR sample, inclination is proposed to change with \fwhb{} and hence we split that sample into ten bins of \fwhb; in the outflow sample, inclination is proposed to change with \dvcm{} and hence we split that sample into ten bins of \dvcm. Each bin is chosen to contain the same number of quasars. The mean SF relations per binned subsample are shown in Figure~\ref{fig:lumoff}, where the bin samples are colour-coded by \fwhb{} and \dvcm{}. As in \citetalias{TWT}, only the SFs belonging to the random walk regime are included in the figure. We fit the $A(\Delta t/t_{\rm th})$ of both individual quasar SFs and of bin sample SFs with a fixed $\gamma_{\rm th}=0.51$ slope to obtain the amplitude offset, \aoff. Using Equations~\ref{eq:tth} and~\ref{eq:A_ttherm} we find a translation from \aoff{} into a luminosity offset, \loff, as: 
\begin{equation}
    \label{eq:lumoff}
    \begin{split}
        \Delta \log A_0 & =\log A_{\rm 0, quasar}-\log A_{\rm 0, uni} \\
            & =\gamma_{\rm th} (\log t_{\rm th, quasar} - \log t_{\rm th, uni}) \\
            & =0.5\gamma_{\rm th} (\log L_{\rm 3000, quasar} - \log L_{\rm 3000, uni}) \\
            & =0.5\gamma_{\rm th} \Delta \log L_{3000}, \\
    \end{split}
\end{equation}
where $A_{\rm 0, uni}$, $t_{\rm th, uni}$, and $L_{\rm 3000, uni}$ are the SF parameters of the mean relation, while $A_{\rm 0, quasar}$, $t_{\rm th, quasar}$, and $L_{\rm 3000, quasar}$ are the SF fit parameters found for the individual quasars or bin samples. To constrain the fit to the random walk regime, we avoid data sections where $\log A \leq -1.4$ or $\log (\Delta t/t_{\rm th}) \leq -0.42$. The \loff{} is calculated for each passband of a quasar SF. Note that this definition makes \loff{} positive when the disc appears overluminous in our assumption and thus the relation is moved to the left of the mean relation in Figure~\ref{fig:lumoff}.

\subsection{Accretion disc luminosity}\label{sec:acc-disc-lum}

We inspect the relation between inclination angle, $i$, and luminosity offset by calculating the theoretical accretion disc luminosity using a steady-state, optically-thick, and geometrically thin accretion disc model, called \texttt{kerrbb} \citep{Li05}. We access the pre-calculated spectral tables through \texttt{XSPEC} 12.12.0 \citep{XSPEC}, packaged in \texttt{Sherpa} \citep{Sherpa}, the modelling and fitting suite of Chandra Interactive Analysis of Observations (CIAO) v4.14. The \texttt{kerrbb} spectra are calculated by ray-tracing from an inclined observer, assuming that the disc radiates locally as a blackbody. All general-relativistic effects such as the Lense-Thirring effect, Doppler boosting, and gravitational redshifting are included. 

As a representative example, we consider a model black hole with $\log(M_{\rm{BH}}/M_{\odot}) = 9$. We set the Eddington ratio to 0.5, radiative efficiency to 10\%, hardening factor to unity, and prescribe zero torque at the inner edge of the disc. Using pre-calculated spectral tables with limb-darkening, we evaluate the monochromatic accretion disc luminosity at 3000\AA~for a range of spins and inclinations. We find only a subtle dependence on input black hole model parameters, except at inclinations close to 90$^\circ$, which are not expected in optically selected samples due to strong extinction from a dusty torus. At this wavelength, we also find no discernible difference in the relative luminosity offsets due to inclination between black hole spins. This remains the case for black hole masses ($M_{\rm BH}$) up to $\log(M_{\rm{AD}}/M_{\odot}) \leq 11$, which encompasses all known black holes with reliable mass measurements. Then, according to the thin disc model \texttt{kerrbb}, the relative accretion disc luminosity at 3000\AA~between inclination angles for type-1 quasars can be considered as a global relation, consistent to within 10\% over three orders of magnitude in $M_{\rm BH}$ from $\log(M_{\rm{AD}}/M_{\odot}) = $ 6 -- 9.

\begin{figure}
\includegraphics[width=\columnwidth]{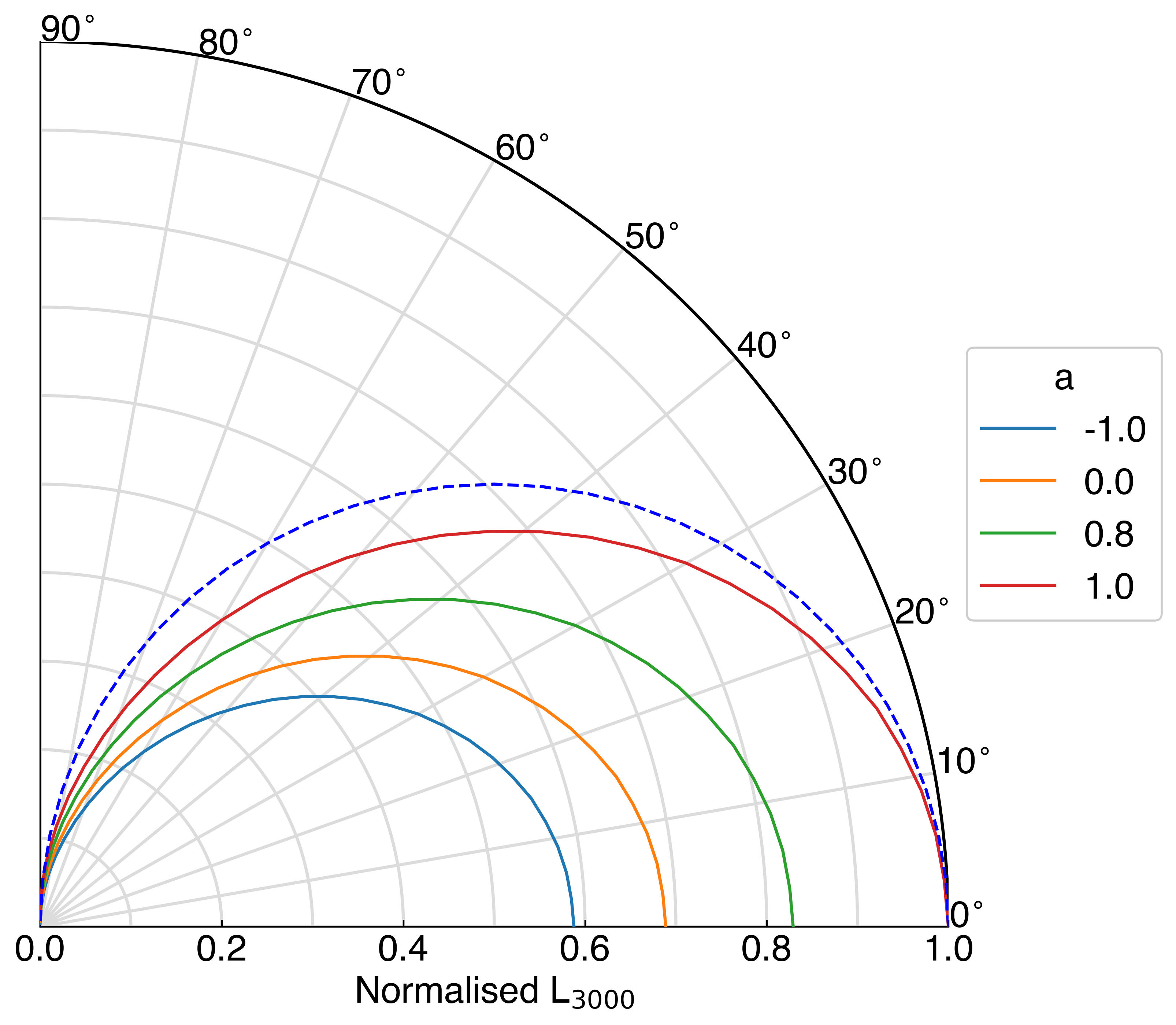}
\caption{Luminosity at 3000\AA~normalised to the maximum luminosity for the full range of inclination angles and black hole spins. The dashed blue line represents the classical $\cos{i}$ luminosity relation. The black hole model parameters are described in Section \ref{sec:acc-disc-lum}, but once normalised to a common value at $i=0$, the shape of all curves is consistent to within 10\%, independent of the assumed black hole model.
(A colour version of this figure is available in the online journal.)
}
\label{fig:acc-disc-lum}
\end{figure}

In Figure~\ref{fig:acc-disc-lum}, we present the accretion disc luminosity at 3000\AA~for a model black hole with the parameters listed above. These curves trace the model flux for all observed $i$ from $0^\circ$ -- $90^\circ$ and individual curves are separated by the black hole spin parameter. The shape of the curves, indicating the relative luminosity offset, are virtually indistinguishable between black hole spins. These luminosity curves are used to estimate the expected luminosity offsets \loff. Refer to the Appendix of \citet{Lai23} for a deeper investigation into the bolometric and monochromatic luminosities at various black hole spins and observed inclinations using the thin disc model.

\section{Models Estimating inclination angle}\label{sec:mod}

We explore the proposed relations between the inclination angle, $i$, and observable quasar properties, which is the \fwhb{} for the rotating BLR sample and the \dvcm{} velocity shift for the outflow sample, with several models.

\subsection{Rotating BLR model for \texorpdfstring{\hbeta}{}}\label{sec:hbmod}

The \hbeta line traces a part of the broad line region (BLR) in a quasar, which has most likely a disc-like geometry \citep[e.g.][]{Panco14}. \citet{MR18} show that the FWHM of broad emission line is anti-correlated with the virial factor due to inclination. Therefore, \fwhb{} can be written \citep{Marz18, Panda19} as:
\begin{equation}
    \label{eq:fwhm}
    {\text \fwhb}=\sqrt{\frac{4(\kappa^2+\sin^2{i})M_{\rm BH}G}{{\text \rblr}}}, 
\end{equation}
where $\kappa$ is the ratio between the isotropic velocity and the Keplerian velocity, \rblr{} is the BLR radius, $M_{\rm BH}$ is the black hole mass, and $G$ is the gravitational constant. We adopt $\kappa=0.1$ from \citet{Marz18} and use the empirical \rblr$-L$ relation,
\begin{equation}
    \label{eq:rl}
    \log({\text \rblr}/ {\rm ld})=K+\alpha\log(\lambda L_{5100} /10^{44} {\rm erg~s}^{-1}),
\end{equation}
to estimate the \rblr{} from the luminosity at 5100\AA~($L_{5100}$). We adopt two sets of [$K$, $\alpha$], [1.527, 0.533] from \citet{Bentz13} and [1.33, 0.41] from \citet{Malik23}. The former is derived from lower accretion rate quasars while the latter includes more sources with higher accretion rate. From \citet{Rak20} we find the three quartile limits for [$\log (\lambda L_{5100}/($erg s$^{-1}$)), $\log (M_{\rm BH}/M_\odot$)] in the full rotating BLR sample as [45.40, 8.7], [45.55, 8.9], and [45.69, 9.1]. Using these values as a first order approach, we can rewrite Equation~\ref{eq:fwhm} into \fwhb{} as a function of $i$. Note that the luminosity range of our sample exceeds the upper limit of quasars used to derive the latest \rblr$-L$ relations, and extrapolation may cause large uncertainties in predicted \rblr{} values. We use results from both relations to indicate the possible range for predicted \fwhb.

\subsection{Outflow model for \texorpdfstring{\civ}{}}\label{sec:symod}

We adopt the kinematic disc-wind model by \citet{Yong16,Yong17,Yong20} for an outflow model with wide opening angles of $15^\circ$ -- $60^\circ$. They sliced the outflow into 16 zones above the disc plane to mimic a stratified BLR. They then used a fixed set of parameters to find the best model to fit the observed properties in quasar spectra from SDSS. As the \civ line is mainly dominated by the wind while the \mgii line is dominated by the rotational component of the accretion disc, they chose two zones close to the pole axis as the \civ emitting region and three zones just above the outer disc plane as the \mgii emitting region \citep[see Figure~15 of][]{Yong20}.

Using the \fwcm{} ratio and \dvcm{} as a \civ blueshift estimate, they proposed a combination of the two correlated variables as an inclination indicator. In their model, pole-on orientations have larger \civ blueshifts and \fwcm{} ratios while edge-on has the opposite behaviour. In Section~\ref{sec:toymod} we confirm with toy models that this behaviour should be qualitatively universal for most outflows. In this study, we use the predicted \dvcm{} blueshift and $i$ values from their model with a $M_{\rm BH}=10^9M_\odot$ and a fixed set of model parameters as described in \citet{Yong20}.

\subsection{Further outflow toy models for \texorpdfstring{\civ}{}}\label{sec:toymod}

To check whether different models make any difference in the general trend between $i$ and \dvcm, we explore a set of simple toy models. For simplicity, we assume that the outflowing gas has no rotational component in this model, which would merely broaden the observed line profile and not shift the \civ line centroid. This is due to the symmetry of the assumed inclined cone, blueshifted and redshifted photons will appear in equal parts due to orbital motion as long as relativistic and radiative transfer effects are ignored. Typical optical depths of the \civ line are unknown \citep{Bald97}. Hence, we consider just two ideal cases: an optically thin case, where photons reach the observer irrespective of their point of origin in the emission volume; and an optically thick case, where only photons from the surface of the emission cone reach the observer. 

We start from the line profile function, $\phi$, in \citet{RH78}, which can be written as
\begin{equation}
    \label{eq:delta}
    \phi(\lambda,  {\text \vlos})=\delta[\lambda-\lambda_0(1+\frac{{\text \vlos}}{c})], 
\end{equation}
where \vlos{} is the line-of-sight (LOS) velocity of the wind and $\lambda_0$ is the rest-frame wavelength of the emission line \civl. Hence, we construct the line profiles using
\begin{equation}
    \label{eq:op_thin}
    L_\lambda \propto \phi(\lambda,  {\text \vlos}) d {\text \vlos}, 
\end{equation}
and
\begin{equation}
    \label{eq:op_thick}
    L_\lambda \propto \phi(\lambda,  {\text \vlos}) d {\text \vlos} dA_{\rm per},
\end{equation}
for optically thin and thick cases, respectively, based on the geometries described below. The $A_{\rm per}$ is the surface area that is perpendicular to the LOS.

\begin{figure}
\includegraphics[width=\columnwidth]{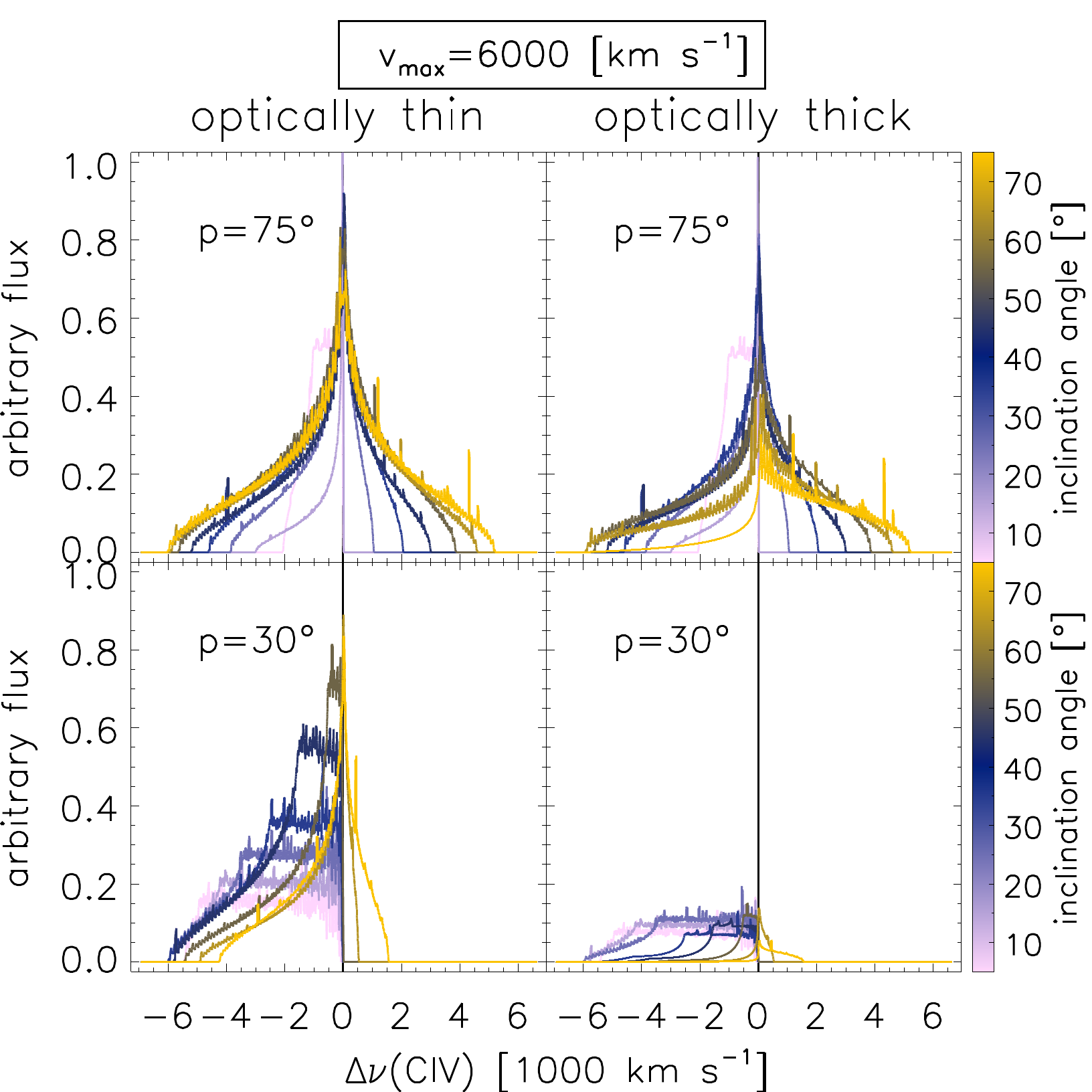}
\caption{Examples of the \civ emission line profile from the \civ outflow toy models with opening angles of $p=30^\circ$ and 75$^\circ$ and for optically thin and thick emission. Modest opening angles are required to produce clearly asymmetric lines with high blueshifts. The profiles are colour-coded by inclination angle.
(A colour version of this figure is available in the online journal.)
}
\label{fig:line_pro}
\end{figure}

The simplest geometry is a perfectly collimated bipolar outflow from the central black hole. We model the outflows with a velocity gradient and use two cases with maximum velocity $v_{\rm max} = 3000$~km~s$^{-1}$ and 6000~km~s$^{-1}$. Based on the assumptions above, we integrate Equation~\ref{eq:delta} over a range of \vlos, which is expected to be $\propto \cos{i}$ in ideal polar outflows. We consider $i$ in the range from $5^\circ$ -- $75^\circ$ to avoid sightlines intersecting a dusty torus as these are not expected in quasar samples. The resulting line profile is a simple step function. From the line profile, we determine the $\lambda_{\rm half}$ where the wavelength bisects the total \civ emission line flux, and translate it into \civ blueshift using 
\begin{equation}
    \label{eq:line_sh}
    {\text \dvciv}=c \times (\lambda_0-\lambda_{\rm half})/\lambda_0,
\end{equation}
where $c$ is the speed of light. Details are shown in Appendix~\ref{app:pole_math}. 

A more realistic model assumes the outflow to have a cone shape with opening angle $p$. We explore $p$ in the range from $15^\circ$ -- $90^\circ$. With these assumptions, we integrate Equations~\ref{eq:op_thin} and~\ref{eq:op_thick} over the \vlos{} given one wind speed of the cone (see Appendix~\ref{app:cone_math} for the resulting equations for the line profiles). We then numerically integrate these line profiles over the two sets of velocity gradients and repeat this over a range in inclination $i$. In Figure~\ref{fig:line_pro}, we show examples of \civ line profiles based on this geometry. It is immediately clear, that highly asymmetric line profiles with significant blueshifts, as observed in many quasars, are only produced by moderate opening angles of the outflow cones. 

\begin{figure}
\includegraphics[width=\columnwidth]{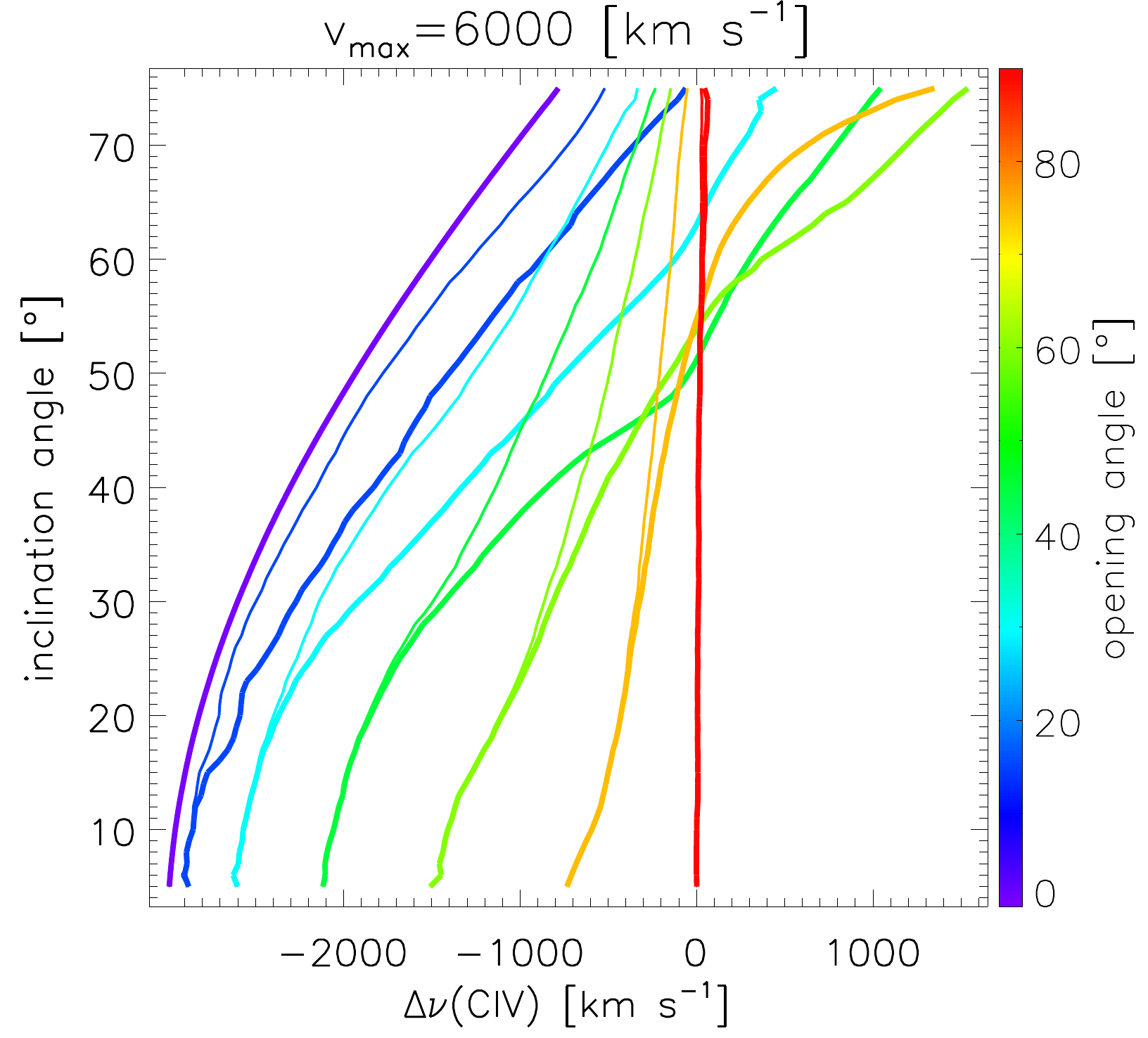}
\caption{Predicted \dvciv{} from the \civ outflow toy models for $v_{\rm max} = 6000$~km~s$^{-1}$, colour-coded by opening angle, for optically thin (thin lines) and thick (thick lines) emission.
(A colour version of this figure is available in the online journal.)
}
\label{fig:dv_incl}
\end{figure}

Finally, we numerically calculate the pivot wavelength, $\lambda_{\rm half}$, and thus the \civ blueshift using Equation~\ref{eq:line_sh}. The summary result of \dvciv{} with all possible ranges of $i$ and $p$ in these models is shown in Figure~\ref{fig:dv_incl} for the case of $v_{\rm max} = 6000$~km~s$^{-1}$. We assume that the \mgii emitting gas is centred on the systemic velocity, so that the derived \civ blueshift corresponds to the \dvcm{} velocity shift. Figure~\ref{fig:dv_incl} shows in more detail that \civ blueshifts are enhanced by lower opening angles of the outflow cones. Irrespective of whether we choose the ideally optically thin or thick model, the trend is always that low-inclination (nearly pole-on) viewing angles produce larger blueshifts than high-inclination ones.

The \civ blueshifts in the \citet{Rak20} catalogue are obtained from the peak wavelengths of the emission lines and not the pivot wavelength that we consider in our models. However, we chose to consider the pivot wavelength due to the simplicity of our models, and expect that peak wavelength and pivot wavelength should be broadly monotonically correlated in observed \civ lines despite noise from complex, real line profiles. Hence, the orientation of the trends in blueshift with inclination should remain the same with either measure even though the scaling may change. In summary, this section has shown that independent of outflow geometry and independent of physical variety in outflow velocities, quasars that are viewed nearly pole-on should indeed show preferentially more blueshifted \civ lines.

\section{Results and discussion}\label{sec:results}

\subsection{Plausible distributions of \texorpdfstring{\civns}{TEXT} blueshifts}\label{sec:toymod_test}

\begin{figure}
\includegraphics[width=\columnwidth]{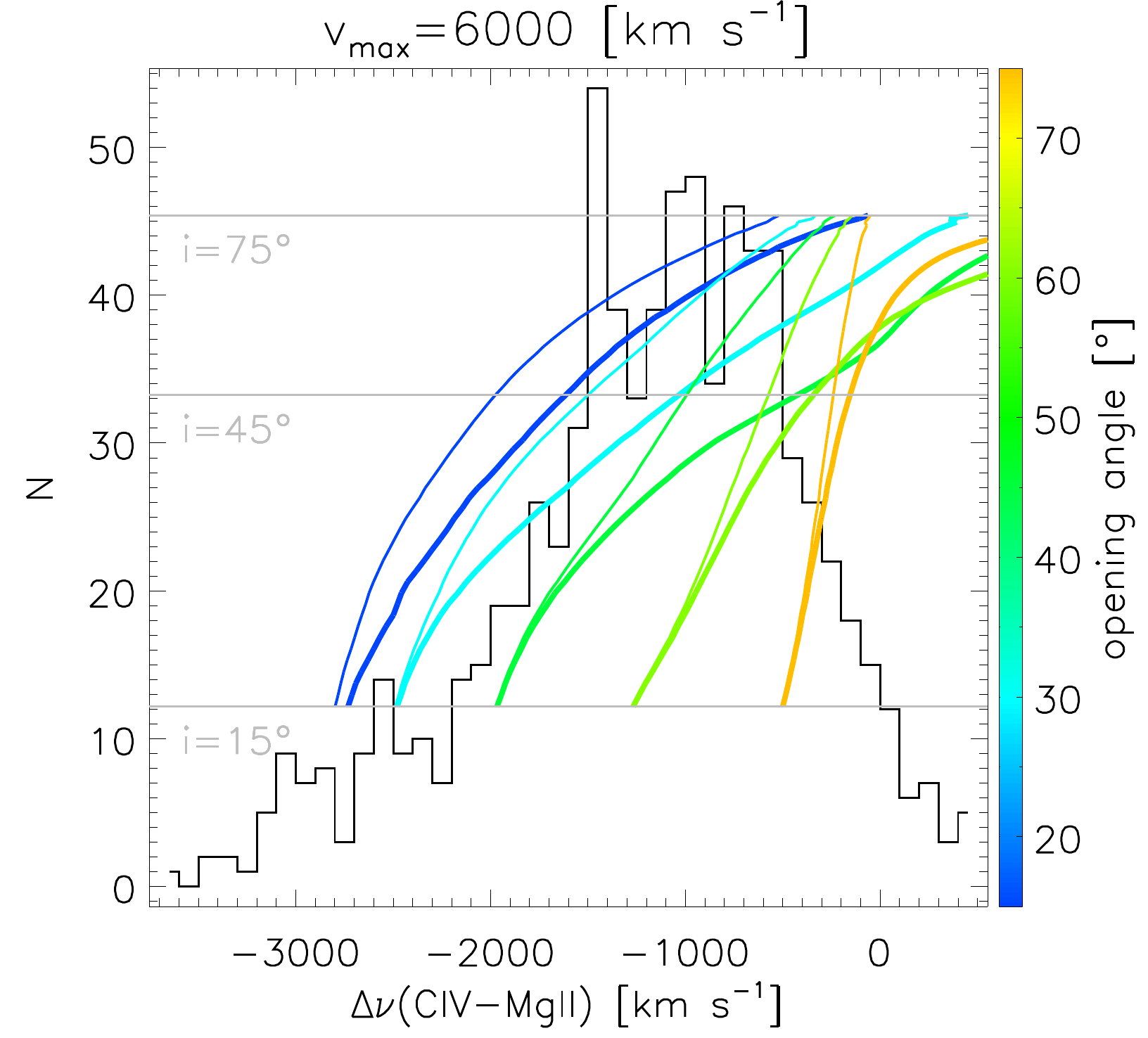}
\caption{Comparison of the \dvcm{} distributions in the outflow sample and the \civ toy models, where \mgii{} is assumed to be centred on the systemic velocity. The predictions are colour-coded by opening angle. Thin and thick lines show the optically thin and thick cases, respectively. As the models are scaled to a common peak $N_{\rm max}$, the mapping $N(i)$ is the same for all toy models; thus, some inclination angles $i$ are shown as horizontal grey lines.
(A colour version of this figure is available in the online journal.)
}
\label{fig:civmod_data}
\end{figure}

We first check whether the toy models in Section~\ref{sec:toymod} are compatible with the outflow sample in terms of the distribution of blueshifts. If the terminal outflow velocity $v_{\rm max}$ is fixed and the inclination angle, $i$, in a quasar sample is randomly distributed, then we expect a distribution of $N(i)\propto \sin{i}$, except for objects missing at high $i$ due to torus obscuration. In Figure~\ref{fig:civmod_data}, the histogram of blueshifts in the outflow sample peaks at \dvcm~$\sim -1000$~km~s$^{-1}$ and decreases toward faster blueshifts. We overlay lines of $\sin{i}$ with $i$ inferred from the \dvciv{} in the toy models with $v_{\rm max}$ = 6000~km~s$^{-1}$, and scale them to roughly match the height of the sample histogram. Note, that changing $v_{\rm max}$ in the model could stretch the $\Delta v$ scale arbitrarily, so the comparison concerns only the qualitative shape of the lines relative to the histogram. The toy models appear broadly consistent with the outflow sample, as a random orientation of the quasars' polar axes predicts large blueshifts to be rare, just as observed.

\subsection{Inclination estimates from emission lines and variability}\label{sec:comp_res}

\begin{figure}
\includegraphics[width=\columnwidth]{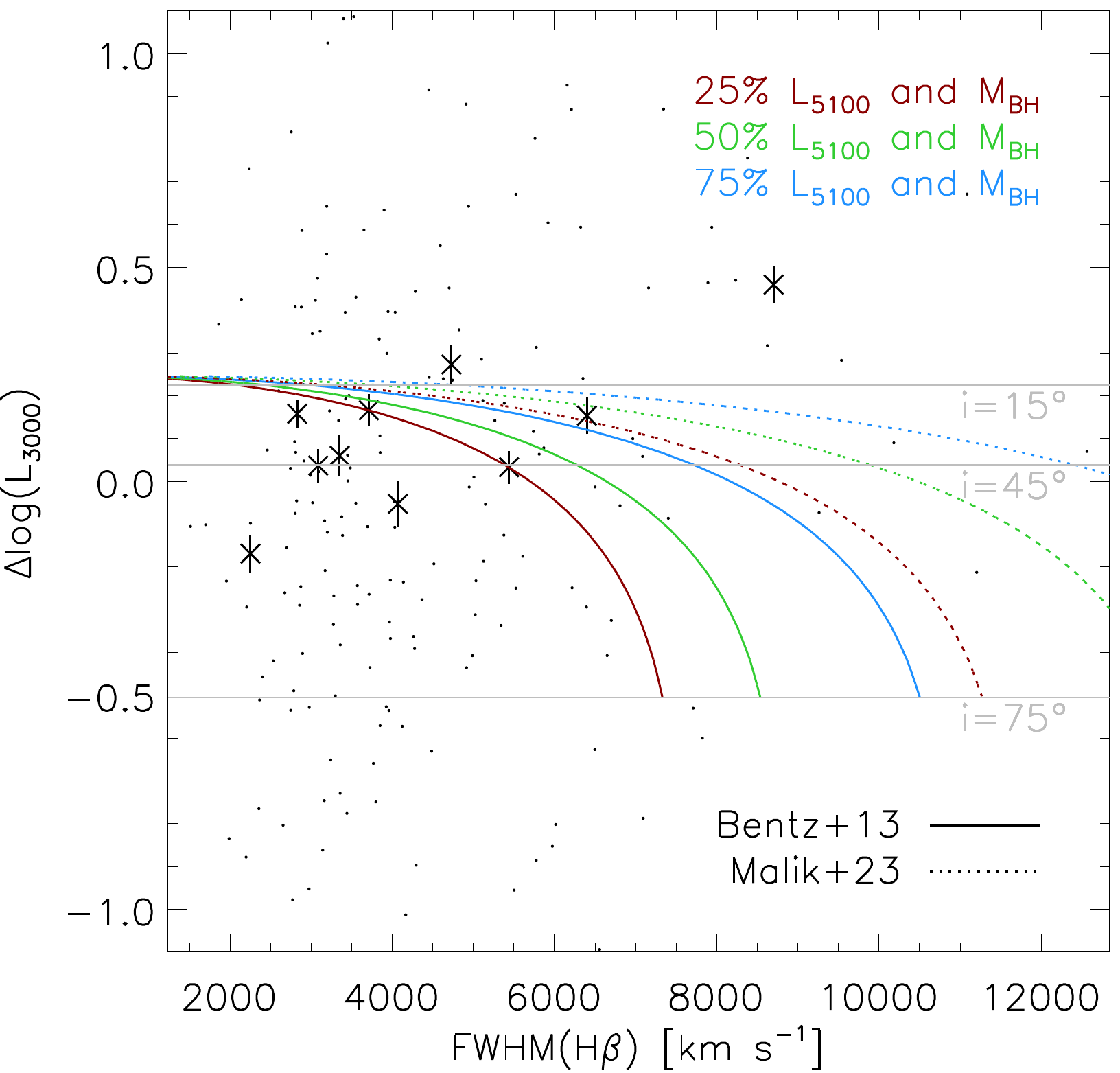}
\caption{Luminosity offset \loff{} versus \fwhb{} in the rotating BLR sample, derived from amplitude offsets \aoff{} in a structure function fit using Equation~\ref{eq:lumoff}. Each dot is an offset for one quasar measured in one passband. The crosses with error bars represent the same quasar groups used for Figure~\ref{fig:lumoff}. Model results from Section~\ref{sec:hbmod} are shown as coloured lines representing percentiles of the $L_{5100}$ and $M_{\rm BH}$ distribution in the full rotating BLR sample. Solid and dotted lines result from two different \rblr$-L$ relations. Horizontal lines indicate the expected \loff{} of selected inclination angles $i$. 
(A colour version of this figure is available in the online journal.)
}
\label{fig:loff_hbmod}
\end{figure}

We test the utility of the variability structure function as an inclination indicator by comparing it with $i$ estimated by \fwhb. Figure~\ref{fig:loff_hbmod} shows the luminosity offset \loff{} versus \fwhb{} in the rotating BLR sample, derived from amplitude offsets \aoff{} in a SF fit using Equation~\ref{eq:lumoff}. Each dot is an offset for one quasar measured in one passband. We split the full sample into ten bins and show their median offsets as crosses and the standard errors of the medians as error bars. These ten bins are the same bin samples that were used in the left panel of Figure~\ref{fig:lumoff}. Using the full sample or the luminosity-complete sample gives consistent results within the noise, so only the results from the full sample are shown. We see that \loff{} tends to increase with broader \fwhb{}, i.e., high-\fwhb{} quasars appear overluminous relative to the mean relation of the quasar variability SF.

In order to compare this observation with the model prediction, we overlay a predicted \loff{} vs. \fwhb{} by combining two relations: (i) the relation between \loff{} and $i$ (Section~\ref{sec:acc-disc-lum}), which is effectively independent of the black hole spin, and (ii) the relation between $i$ and \fwhb{} (Section~\ref{sec:hbmod}), which depends on $L_{5100}$ and $M_{\rm BH}$. The \loff{} for the model predictions have been normalised by shifting them to an average value of 0. In the model, the predicted \loff{} decrease towards higher $i$ and thus broader \fwhb, while the observed result has exactly the opposite trend. 

The models include a relation between the size of the BLR and the luminosity of the accretion disc, for which we employ two versions, from \citet{Bentz13} and \citet{Malik23}. As we can see in the figure, these versions merely rescale the \fwhb{} observed for a given luminosity and inclination angle, but do not alter the basic shape of the model curves.

\begin{figure*}
\includegraphics[width=\textwidth]{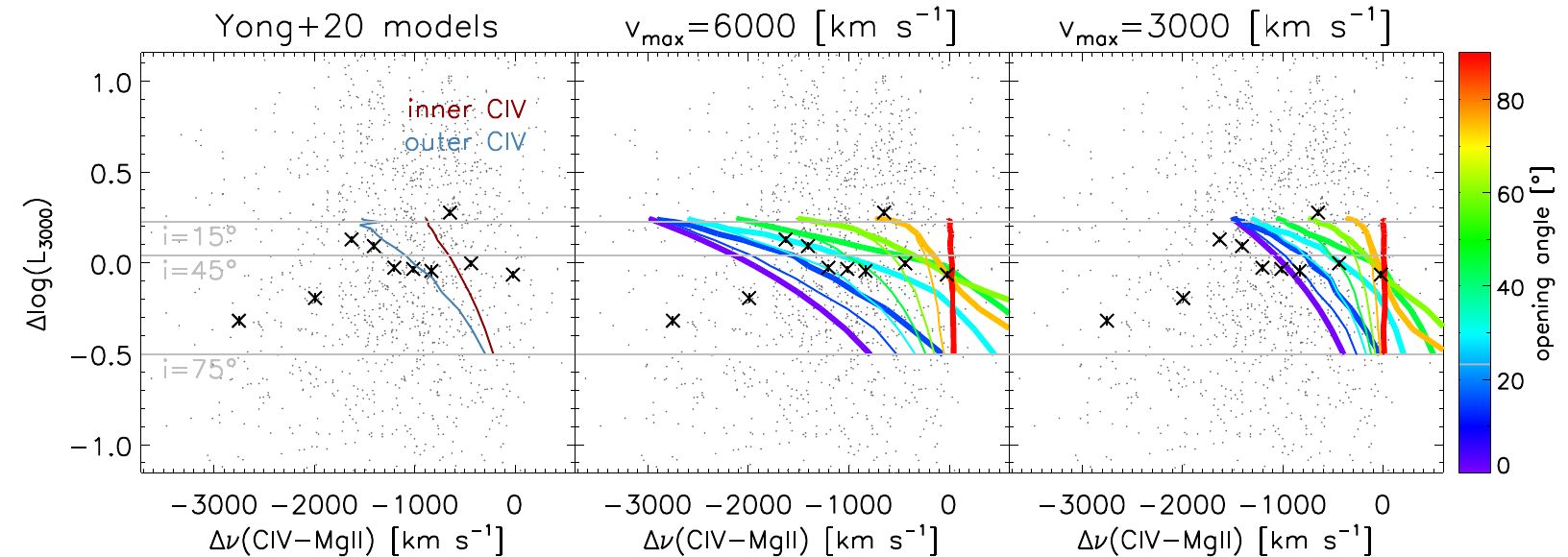}
\caption{Luminosity offset \loff{} versus \dvcm{} in the outflow sample, in analogy to Figure~\ref{fig:loff_hbmod}, see caption for details. Here, the left panel shows the model from Section~\ref{sec:symod}. The centre and right panels show the toy models from Section~\ref{sec:toymod} with two examples for the terminal outflow speed $v_{\rm max}$ and colour-coded by opening angle $p$. Thin and thick lines show the optically thin and thick cases, respectively. 
(A colour version of this figure is available in the online journal.)
}
\label{fig:loff_civmod}
\end{figure*}

We also compare the inclination inferred by variability and the inclination estimated from \dvcm, using the models of \citet{Yong20} as well as our toy models. Figure~\ref{fig:loff_civmod} shows the luminosity offset \loff{} versus \dvcm{} in the outflow sample, derived from amplitude offsets in analogy to Figure~\ref{fig:loff_hbmod}. Here, the crosses and error bars represent ten bins in \dvcm{} with the same bin samples that were used in the right panel of Figure~\ref{fig:lumoff}. The observations show that \loff{} increases with decreasing \civ blueshift, although there are hints of a flattening at $\Delta v > -1800$~km~s$^{-1}$. 

We compare this with model predictions using the relations between $i$ and \dvcm{} from Section~\ref{sec:symod} and~\ref{sec:toymod}. No matter which geometric model we adopt, the predicted \loff{} decrease with higher $i$ thus decreasing \civ blueshift, which is again the opposite of the observed result. Quasars with high-\civ blueshift appear underluminous even though the opposite is expected from anisotropic disc emission and projected radial outflow velocities. We note, that this result is consistent with \citet{Yu22} who also find that quasars with higher \civ blueshift have lower variability amplitudes, although they make no assumptions on why that might be.

In summary, we note that although the use of \fwhb{} and \civ blueshift as orientation indicators is not firmly established, the trends we find for both of these two emission lines are consistent. The SF analysis suggests that quasar discs which are presumably viewed pole-on appear underluminous, while those presumed to be viewed at high inclination appear overluminous. This finding contradicts general expectations from the anisotropy of the disc emission, and indeed contradicts consistently for both indicators of viewing angle.

\subsection{Possible reasons for the unexpected result}\label{sec:dis}

Figures~\ref{fig:loff_hbmod} and \ref{fig:loff_civmod} indicate that there is indeed a trend of variability with respect to the \fwhb{} and the \civ blueshift, but it goes in the opposite direction of the trends predicted by the models, which combine calculations of the anisotropy in the disc emission with models predicting emission line properties. Either there are flaws predicting $i$ from \fwhb{} and \civ blueshift, which conspire to produce a similarly wrong estimate of inclination from both line diagnostics. Or there are issues with the brightness of the accretion disc expected as a function of inclination. A third option is that viewing angle affects the variability signal we observe. And a fourth option is that low-\fwhb{} and high-\civ blueshift objects are physically different. In the following, we discuss these four options.

\subsubsection{Are the line diagnostics wrong (inverted)?}
\label{sec:wrongline}

First, we consider the option of the inclination diagnostics from the emission lines being wrong and that of the disc luminosity being trustworthy. If the lines are still inclination indicators, it is possible that the \civ emitting gas could be in an equatorial configuration and display a blueshift due to partial obscuration when viewed near edge-on \citep{Rich02}, although later studies oppose this hypothesis and proposed a disc wind model instead \citep[e.g.][]{Lei04, Rich11}. 

Several studies showed that the \civ blueshift is anti-correlated with the \fwhb{} among quasars \citep[e.g.][]{SL12, Coat16, Yong20}. Combining Figures~\ref{fig:loff_hbmod} and~\ref{fig:loff_civmod}, we see that the anti-correlation holds as in negative luminosity offsets correlate with both low-\fwhb{} and high-\civ blueshift. However, it seems rather difficult to explain how the \fwhb{} may be maximised in a pole-on orientation, where the discs appear overluminous, while \hbeta is a symmetric line centred on the systemic redshift. Also, \citet{MR18} present strong evidence for disc-like motions by showing that the difference between black-hole mass estimates from fitting accretion-disc spectral energy distributions (SEDs) and those from virial estimates based on line widths depends on line FWHM, and that the disagreement can be resolved if FWHM depends on inclination as expected from disk-like orbital motions.

\begin{figure}
\includegraphics[width=\columnwidth]{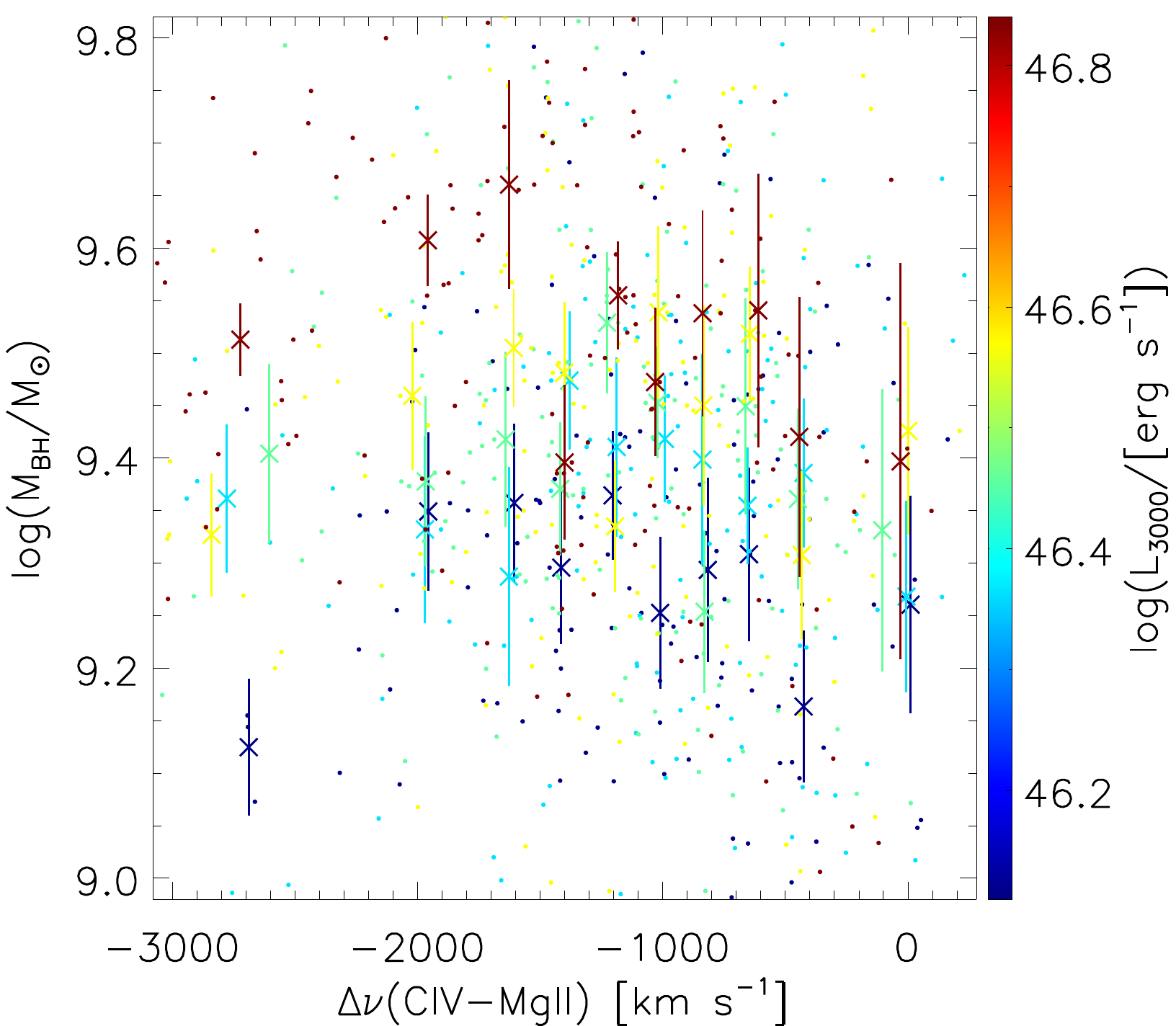}
\caption{$\log(M_{\rm{BH}})$ versus \dvcm{} in the outflow sample.
Each dot represents a quasar. We group them according to their $\log L_{3000}$ and show the median and standard error of the median of $\log(M_{\rm{BH}})$ in cross and error bar, respectively. (A colour version of this figure is available in the online journal.)
}
\label{fig:civ_mbh}
\end{figure}

\subsubsection{Is the disc luminosity diagnostic wrong?}
\label{sec:wronglum}

Second, we consider the option of the inclination diagnostics from the anisotropic disc emission being wrong and that of the emission lines being trustworthy. So far, we assumed that the observed \loff{} is caused by inclination, and based on thin-disc models the effect seems inevitable. However, it is possible that dust extinction plays an additional role. Dust is clearly present in the torus surrounding the accretion disc and BLR \citep{UP95} and may additionally be present in the BLR itself \citep{CH11}.

The \loff{} between highly blueshifted and less blueshifted \civ lines is about $-0.25$~dex, while about $+0.25$~dex (in the opposite direction) is expected from the disc inclination effects. This means that a 0.5~dex difference in disc luminosity needs to be explained by dust extinction being higher for the pole-on sample than for the high-inclination sample by about $\Delta A_{3000} \simeq 1.25$~mag. This corresponds to an additional reddening of $\Delta E(B-V)\simeq 0.25$ using a dust extinction law $A_{3000}/A_{\rm V}=1.86$ from \citet{Gal10} and Small Magellanic Cloud (SMC)-like $R_{\rm V}=2.7$ \citep{Bouch85}. While an elevated dust extinction around the polar axis of quasar accretion discs is not strictly ruled out, this is not at all a scenario expected by unified models. This point could be studied further by analysing SED differences between high- and low-\civ blueshift quasars in the outflow sample \citep[e.g.][]{Temp21}, or between low- and high-\fwhb{} samples.

\subsubsection{Could the observed variability depend on viewing angle?}
\label{sec:wrongvar}

The third option for resolving this conundrum is that the variability observed in the light curve itself is altered by the viewing angle. In this case, there is still a bias in observed luminosity from the anisotropic disc emission, which would move the observed variability SF for highly inclined discs to the right of the mean relation in Figure~\ref{fig:lumoff}, such that from this first step alone the timescale will appear too long at a given amplitude or the amplitude will appear too low at a given timescale. Of course, the SF analysis shows high-\fwhb{} and low-\civ blueshift objects, which are presumed to be the more inclined discs, with amplitudes above the mean relation. Thus, the shift right from the underluminous appearance of the disc setting the wrong timescale units would need to be more than compensated by another effect. In principle, we can see two possibilities:

\begin{enumerate}
\item An increase of the observed variability amplitude may shift the SF above the mean relation, which may be caused by gravitational lensing as the black hole causes periodic magnification of disc inhomogeneities on an orbital timescale, especially in the inner parts of the accretion disc \citep{Fuk03,BD05}.
\item Partial dust obscuration of the disc may shift the SF back to the left; if just under half of the disc is obscured by dust from a flaring torus, then the hotter, inner disc with its shorter \tth{} may be fully visible, while the cooler outer disc with its longer timescales may add only around half of its emission to the LC, shifting the mean timescale for variability shorter.  
\end{enumerate}

Calculations of such effects are beyond the scope of this paper, but we note that partial dust obscuration would affect the integrated disc SED as well, as its cooler parts would end up underrepresented.

\subsubsection{Are the orbital timescales wrong in high-\texorpdfstring{\civns}{TEXT} blueshift objects?}\label{sec:wrongmodel}

If the line properties are not primarily caused by orientation, then the high-\civ blueshift and/or low-\fwhb{} samples may be dominated by quasar discs, whose orbital timescale differs from that of the majority of quasars at the same observed luminosity. This means that either the true luminosity is not recognised, perhaps due to dust extinction, or that the temperature profile of the discs is different and implies a different orbital timescale at the disc radii which emit the measured variable brightness.

\citet{Lei04} perform a photoionisation calculation to demonstrate that a quasar with steep optical-to-Xray slope (deficient X-ray photons) allows the gas above the disc to accelerate before it gets ionized, forming a wind with high-\civ blueshift. Indeed, \citet{Rich11} find evidence that quasars with steeper optical-to-Xray slopes have higher \civ blueshifts. Several studies also show that \civ blueshift is correlated with Eddington ratio \citep[e.g.][]{Rank20}, which is anti-correlated with variability amplitude \citep{Yu22}.

\citet{Rich11} further suggest that high-\civ blueshift objects have stronger-than-normal disc winds with embedded dust, which could absorb part of the disc continuum and render the discs underluminous relative to those with weaker outflows at the same viewing angle. If we assume no mean difference in orientation between high- and low-\civ blueshift samples, a $\Delta \log L_{3000} = 0.26$~dex suggests that the outflow itself adds about $\Delta A_{3000} \simeq 0.65$~mag of dust extinction. \citet{Temp21} show that objects with high-\civ blueshift have stronger thermal emission from hot dust, with temperatures close to the dust sublimation limit. This suggests that the extra dust emission in windy quasars is not powered by star formation but correlated with the outflow from the nucleus, and given its high temperature probably close to the accretion disc. Perhaps it is dust in the wind that causes both the absorption we may see manifest as our underluminous discs and the IR emission seen in \citet{Temp21}. Using the \cite{Gal10} extinction curve, we find that the required level of dust extinction makes the power law continuum, $f_\nu\propto\nu^{\alpha_\nu}$, redder by $\Delta \alpha_\nu \approx 0.5$. The \citet{Temp21} comparison of rest-frame UV SEDs between windy and normal quasars shows no such effect, ruling out the dust extinction scenario unless dust in the intense radiation field of the inner AGN has a flat extinction curve in the UV. 

Finally, we consider that the accretion discs in the high-\civ blueshift sample have a different temperature profile from the standard thin-disc approximation. This is the case e.g. in slim discs \citep{Ab88,Sad11}, which are expected to form at larger Eddington ratios. In this scenario, the light observed through the passbands would originate from different radii than in the thin-disc approximation and thus a different orbital timescale could offset the variability SF from the global mean. If we increase our estimate of the orbital timescale in high-\civ blueshift objects by $\sim 0.13$~dex, then their offset from the majority of the sample disappears. 

For this point, we have calculated the emission profile of non-relativistic discs for the median redshift and black-hole mass of the outflow sample. We varied the exponent of the temperature profile and in correlation also the disc scale to keep the monochromatic luminosity fixed at its median value. Integrating over the ATLAS cyan and orange passbands, we derive the mean orbital timescale as a function of radial temperature exponent. We find that a timescale increased by 0.13~dex could be caused by a flatter temperature profile of $T \propto R^{-0.64}$, which stretches the tail of emission at a given wavelength to larger radii in comparison to the thin-disc profile of $T \propto R^{-3/4}$. However, we note that such a flatter profile will also make the slope of the power law disc continuum redder by $\Delta \alpha_\nu \sim 0.45$, which is similarly in disagreement with the \citet{Temp21} SED comparison. 

For completeness, we note that since slim discs are expected to be related to high Eddington ratios, we check our outflow sample for any trends of that with respect to \civ blueshift. Figure~\ref{fig:civ_mbh} shows the black holes mass $M_{\rm BH}$ measured with the virial single-epoch method \citep{MJ02} from the \mgii line, where we adopt the values from the \citet{Rak20} catalogue. At fixed luminosity we find no trend. We note that \citet{Temp23} found a preference of high-\civ blueshifts to go with large black-hole masses and high luminosity, and we suspect that the difference is because our sample is limited to the brightest quasars from the start.

\section{Conclusions}\label{sec:conclusion}

In this study, we examine the hypothesis of whether quasar variability can be used to determine the inclination angle, $i$, of the accretion disc in active galactic nuclei. We adopt the universal mean relation describing the quasar variability structure function in the random walk regime from \citetalias{TWT}, which includes a dependence on the luminosity of the accretion disc. We test the hypothesis with two sub-samples of the \citetalias{TWT} sample, a rotating broad-line region sample and an outflow sample. Assuming that the structure function of each quasar is offset from the mean relation by the viewing angle affecting the measured luminosity, we compare it with two spectral inclination indicators proposed earlier, \fwhb{} for the rotating BLR sample and \dvcm{} for the outflow sample. We calculate the theoretical luminosity offset as a function of inclination using the \texttt{kerrbb} model, and predict expected relations with the spectral indicators. We find that both the rotating BLR sample and the outflow sample show a trend of the luminosity offset with respect to the \fwhb{} and \dvcm{}, respectively, that is the opposite of the predicted trend. 

The result suggests that quasars with highly blueshifted \civ lines differ from the majority of the quasar sample certainly by more than just inclination effects. We speculate that they may have discs with a flatter temperature structure or dusty outflows rendering the accretion disc underluminous, but we offer no conclusive picture. The dust scenario would only be consistent with quasar SEDs if the UV extinction curves of hot dust were unusually flat. Also, the explanation for the modestly significant trend with \fwhb{} is unclear. Future work may require not only light curves of many more quasars, but extensive data on quasar SEDs to unpick the observed subtle trend residuals in quasar variability.

\section*{Acknowledgements}

We thank an anonymous referee for suggestions improving the manuscript. JJT was supported by the Taiwan Australian National University PhD scholarship, the Australian Research Council (ARC) through Discovery Project DP190100252, the National Taiwan University research grant NTU-112L7302, the Ministry of Education, Taiwan Yushan Young Scholar grant NTU-111V1007-2, and also acknowledges support by the Institute of Astronomy and Astrophysics, Academia Sinica (ASIAA). JT has been funded in part by the Stromlo Distinguished Visitor Program at RSAA. We thank Youichi Ohyama for useful discussion and Christopher A. Onken for suggestions improving the manuscript. This research has made use of \textsc{idl} and \textsc{mathematica}.

This work uses data from the University of Hawaii's ATLAS project, funded through NASA grants NN12AR55G, 80NSSC18K0284, and 80NSSC18K1575, with contributions from the Queen's University Belfast, STScI, the South African Astronomical Observatory, and the Millennium Institute of Astrophysics, Chile.

This work has made use of SDSS spectroscopic data. Funding for the Sloan Digital Sky Survey IV has been provided by the Alfred P. Sloan Foundation, the U.S. Department of Energy Office of Science, and the Participating Institutions. SDSS-IV acknowledges support and resources from the Center for High Performance Computing  at the University of Utah. The SDSS website is \href{http://www.sdss.org}{http://www.sdss.org}. SDSS-IV is managed by the Astrophysical Research Consortium for the Participating Institutions of the SDSS Collaboration including the Brazilian Participation Group, the Carnegie Institution for Science, Carnegie Mellon University, Center for Astrophysics | Harvard \& Smithsonian, the Chilean Participation Group, the French Participation Group, Instituto de Astrof\'isica de Canarias, The Johns Hopkins University, Kavli Institute for the Physics and Mathematics of the Universe (IPMU) / University of Tokyo, the Korean Participation Group, Lawrence Berkeley National Laboratory, Leibniz Institut f\"ur Astrophysik Potsdam (AIP),  Max-Planck-Institut f\"ur Astronomie (MPIA Heidelberg), Max-Planck-Institut f\"ur Astrophysik (MPA Garching), Max-Planck-Institut f\"ur Extraterrestrische Physik (MPE), National Astronomical Observatories of China, New Mexico State University, New York University, University of Notre Dame, Observat\'ario Nacional / MCTI, The Ohio State University, Pennsylvania State University, Shanghai Astronomical Observatory, United Kingdom Participation Group, Universidad Nacional Aut\'onoma de M\'exico, University of Arizona, University of Colorado Boulder, University of Oxford, University of Portsmouth, University of Utah, University of Virginia, University of Washington, University of Wisconsin, Vanderbilt University, and Yale University.

This work has made use of data from the European Space Agency (ESA) mission Gaia (\href{https://www.cosmos.esa.int/gaia}{https://www.cosmos.esa.int/gaia}), processed by the Gaia Data Processing and Analysis Consortium (DPAC, \href{ https://www.cosmos.esa.int/web/gaia/dpac/consortium}{ https://www.cosmos.esa.int/web/gaia/dpac/consortium}). Funding for the DPAC has been provided by national institutions, in particular the institutions participating in the Gaia Multilateral Agreement.

\section*{Data Availability}
The data underlying this article will be shared on reasonable request to the corresponding author. 





\begin{landscape}
\begin{appendix}
\section{Exact solution of the bipolar outflow model}\label{app:pole_math}

In the bipolar outflow, the opening angle, $p = 0$. The velocity along the line-of-sight (LOS), \vlos, is 
\begin{equation}
    \label{eq:pole_vlos}
    {\text \vlos}=v_{\rm local }\cos{i},
\end{equation}
where \vloc{} is the wind speed of the outflow local element.
In the following, we define \bloc$=\frac{{\text \vloc}}{c}$.

If we substitute Equation~\ref{eq:pole_vlos} into Equation~\ref{eq:delta}, we get the line profile as
\begin{equation}
    \label{eq:pole_lum}
    \phi(\lambda, {\text \bloc})=\delta[\frac{1}{\lambda}-\frac{1}{\lambda_0}(1+{\text \bloc}\cos{i})].
\end{equation}
We assume a velocity gradient in the slow and fast outflows, which are \vloc{} between [30, 3000]~km~s$^{-1}$ and between [60, 6000]~km~s$^{-1}$, respectively. Each of the outflows consists of 100 speed elements, which are evenly distributed among the above \vloc{} range. We integrate Equation~\ref{eq:pole_lum} over the \bloc{} range [\bmin, \bmax] and get
\begin{flalign}
    L_\nu \propto \int_{\text \bmin}^{\text \bmax} \delta[\frac{1}{\lambda}-\frac{1}{\lambda_0}(1+{\text \bloc}\cos{i})] d{\text \bloc}
    =\left\{
    \begin{matrix}
    \label{eq:pole_int}
        \frac{\lambda_0}{\cos{i}} & \text{ for } \frac{\lambda_0}{1+\beta_{\rm local, max}\cos{i}}<\lambda<\frac{\lambda_0}{1+\beta_{\rm local, min}\cos{i}}\\
        0 & \text{ for $\lambda$ everywhere else.}
    \end{matrix}
    \right.
\end{flalign}
Since Equation~\ref{eq:pole_int} is a step function, it is obvious that the $\lambda_{\rm half}$ is the middle between the two boundaries. If we substitute it into Equation~\ref{eq:line_sh}, we will get 
\begin{equation}
    \label{eq:pole_sh}
    {\text \dvciv}=c \times \frac{0.5({\text \bmax}+{\text \bmin})\cos{i}+{\text \bmax}{\text \bmin}\cos^2{i}}{1+({\text \bmax}+{\text \bmin})\cos{i}+{\text \bmax}{\text \bmin}\cos^2{i}} \approx c\times0.5({\text \bmax}+{\text \bmin})\cos{i}.
\end{equation}

\section{Exact solution of the cone outflow model}\label{app:cone_math}

For the cone outflow, the opening angle, $p > 0$. If we adopt the spherical coordinate and define an angle, $\theta$, between each outflow infinitesimal beam element and zenith, the \vlos{} is 
\begin{equation}
    \label{eq:cone_vlos}
    {\text \vlos}={\text \vloc}(\cos{p}\cos{i}-\sin{p}\sin{i}\cos{\theta}),
\end{equation}
where \vloc{} is the wind speed of the outflow local element. Again, we define \bloc$=\frac{{\text \vloc}}{c}$.

In the optically thin case, substituting Equation~\ref{eq:cone_vlos} into Equation~\ref{eq:op_thin} gives 
\begin{flalign}
    \nonumber
    L_\nu &\propto \int^{\pi}_0 \delta[\frac{1}{\lambda}-\frac{1}{\lambda_0}(1+{\text \bloc}(\cos{p}\cos{i}-\sin{p}\sin{i}\cos{\theta}))] d\theta \\ 
    &=\left\{
    \begin{matrix}
    \label{eq:thin}
        \frac{\lambda_0\lambda}{\sqrt{-\lambda_0^2+2(1+{\text \bloc}\cos{p}\cos{i})\lambda_0\lambda-[1+2{\text \bloc}\cos{p}\cos{i}+{\text \bloc}^2(\cos^2{p}-\sin^2{i})]\lambda^2}}
        & \text{ for } \frac{\lambda_0}{1+{\text \bloc}\cos{(p-i)}}<\lambda<\frac{\lambda_0}{1+{\text \bloc}\cos{(p+i)}}\\
        0 & \text{ for $\lambda$ everywhere else.}
    \end{matrix}
    \right.
\end{flalign}
Note that this is applicable for all possible $p$ and $i$.

In the optically thick case, we simplify the calculation by defining a new angle, $\theta'$, between direction towards the observer and each outflow infinitesimal beam element. The relation between $\theta'$ and $\theta$ is $\cos{\theta'}=\cos{p}\cos{i}-\sin{p}\sin{i}\cos{\theta}$. We need to consider the relation between $p$ and $i$. When the observer is viewing the inside of the cone outflow, i.e., $p \geq i$, substituting Equation~\ref{eq:cone_vlos} into Equation~\ref{eq:op_thick} and changing the variable from $\theta$ to $\theta'$ in the integration gives
\begin{flalign}
    \nonumber
    L_{\nu, \rm inner} &\propto \int^{p-i}_{p+i} \frac{\cos{p}\cos{\theta'}-\cos{i}}{\sqrt{\sin^2{i}\sin^2{\theta'}-(\cos{p}-\cos{i}\cos{\theta'})^2}} \delta[\frac{1}{\lambda}-\frac{1}{\lambda_0}(1+{\text \bloc}\cos{\theta'})] d\theta' \\
    &=\left\{
    \begin{matrix}
    \label{eq:thick_in}
        \frac{({\text \bloc}\cos{i}-\cos{p}\lambda_0^2)\lambda+\cos{p}\lambda_0\lambda^2}{\sqrt{[-\lambda_0^2+2\lambda_0\lambda-(1-{\text \bloc}^2)\lambda^2]\{-\lambda_0^2+2(1+{\text \bloc}\cos{p}\cos{i})\lambda_0\lambda-[1+2{\text \bloc}\cos{p}\cos{i}+{\text \bloc}^2(\cos^2{p}-\sin^2{i})]\lambda^2\}}}
        & \text{ for } \frac{\lambda_0}{1+{\text \bloc}\cos{(p-i)}}<\lambda<\frac{\lambda_0}{1+{\text \bloc}\cos{(p+i)}}\\
        0 & \text{ for $\lambda$ everywhere else.}
    \end{matrix}
    \right.
\end{flalign}

\clearpage

When the observer is viewing outside of the cone outflow, i.e., the case of $p < i$, the observer can see the backside and only part of the inner side of the cone. The farther half of the inner side is blocked by the backside of the cone. Therefore, we should consider an additional subtraction term than Equation~\ref{eq:thick_in}, which is 
\begin{flalign}
    \nonumber
    L_{\nu, \rm subtract} &\propto \int^{\cos^{-1}{(\cos{p}\cos{i}})}_{p+i} \frac{\cos{p}\cos{\theta'}-\cos{i}}{\sin{\theta'}}\sqrt{ \frac{1-(2\cos{p}\cos{i}-\cos{\theta'})^2}{\sin^2{i}\sin^2{\theta'}-(\cos{p}-\cos{i}\cos{\theta'})^2}} \delta[\frac{1}{\lambda}-\frac{1}{\lambda_0}(1+{\text \bloc}\cos{\theta'})] d\theta' \\
    &=\left\{
    \begin{matrix}
    \label{eq:thick_bk}
        \frac{({\text \bloc}\cos{i}-\cos{p}\lambda_0^2)\lambda+\cos{p}\lambda_0\lambda^2}{-\lambda_0^2+2\lambda_0\lambda-(1-{\text \bloc}^2)\lambda^2}{\sqrt{\frac{\lambda_0^2-2(1+2{\text \bloc}\cos{p}\cos{i})\lambda_0\lambda+[1+4{\text \bloc}\cos{p}\cos{i}+{\text \bloc}^2(-1+4\cos^2{p}\cos^2{i})]\lambda^2}{-\lambda_0^2+2(1+{\text \bloc}\cos{p}\cos{i})\lambda_0\lambda-[1+2{\text \bloc}\cos{p}\cos{i}+{\text \bloc}^2(\cos^2{p}-\sin^2{i})]\lambda^2}}}
        & \text{ for } \frac{\lambda_0}{1+{\text \bloc}\cos{p}\cos{i}}<\lambda<\frac{\lambda_0}{1+{\text \bloc}\cos{(p+i)}}\\
        0 & \text{ for $\lambda$ everywhere else.}
    \end{matrix}
    \right.
\end{flalign}
In the end, the line profile is $L_{\nu, \rm inner}-L_{\nu, \rm subtract}$ for the case of $p < i$. Those monochromatic luminosity that are below zero after the subtraction are treated as zero in the following $\lambda_{\rm half}$ calculation. We assume the same velocity gradients as the polar outflow (Appendix~\ref{app:pole_math}), and numerically integrate above line profiles over the \bloc{} range for the fast ($v_{\rm max} = 6000$~km~s$^{-1}$) and slow ($v_{\rm max} = 3000$~km~s$^{-1}$) outflows. For simplicity, we assume that different velocity components (\bloc) are blocking each other in the optically thick case. Only the cone from the same \bloc{} will block itself.

\end{appendix}


\bsp	
\label{lastpage}
\end{landscape}
\end{document}